\documentclass[twoside,twocolumn,9pt]{article}
\usepackage[super,sort&compress,comma]{natbib} 
\usepackage[version=3]{mhchem}
\usepackage[left=1.5cm, right=1.5cm, top=1.785cm, bottom=2.0cm]{geometry}
\usepackage{balance}
\usepackage{newtxtext}
\usepackage{sectsty}
\usepackage{graphicx}
\usepackage{lastpage}
\usepackage[format=plain,justification=justified,singlelinecheck=false,font={stretch=1.125,small,sf},labelfont=bf,labelsep=space]{caption}
\usepackage{float}
\usepackage{fancyhdr}
\usepackage{fnpos}
\usepackage[english]{babel}
\addto{\captionsenglish}{%
  
}
\usepackage{array}
\usepackage{droidsans}
\usepackage{charter}
\usepackage[T1]{fontenc}
\usepackage[usenames,dvipsnames]{xcolor}
\usepackage{setspace}
\usepackage[compact]{titlesec}
\usepackage{hyperref}
\usepackage{amsmath}
\usepackage{mathtools}
\usepackage{amssymb}
\usepackage{grffile}
\usepackage{float}
\usepackage{gensymb}
\usepackage{dcolumn}
\usepackage{bm}
\usepackage{multirow}
\usepackage{soul}
\usepackage{caption}
\usepackage{upgreek}
\usepackage{placeins}

\usepackage{epstopdf}

\definecolor{cream}{RGB}{222,217,201}

\usepackage{epstopdf}

\definecolor{cream}{RGB}{222,217,201}

\begin{document}

\pagestyle{fancy}
\thispagestyle{plain}
\fancypagestyle{plain}{
\renewcommand{\headrulewidth}{0pt}
}

\makeFNbottom
\makeatletter
\renewcommand\LARGE{\@setfontsize\LARGE{15pt}{17}}
\renewcommand\Large{\@setfontsize\Large{12pt}{14}}
\renewcommand\large{\@setfontsize\large{10pt}{12}}
\renewcommand\footnotesize{\@setfontsize\footnotesize{7pt}{10}}
\makeatother

\renewcommand{\thefootnote}{\fnsymbol{footnote}}
\renewcommand\footnoterule{\vspace*{1pt}%
\color{cream}\hrule width 3.5in height 0.4pt \color{black}\vspace*{5pt}} 
\setcounter{secnumdepth}{5}

\makeatletter 
\renewcommand\@biblabel[1]{#1}            
\renewcommand\@makefntext[1]%
{\noindent\makebox[0pt][r]{\@thefnmark\,}#1}
\makeatother 
\renewcommand{\figurename}{\small{Fig.}~}
\sectionfont{\sffamily\Large}
\subsectionfont{\normalsize}
\subsubsectionfont{\bf}
\setstretch{1.125} 
\setlength{\skip\footins}{0.8cm}
\setlength{\footnotesep}{0.25cm}
\setlength{\jot}{10pt}
\titlespacing*{\section}{0pt}{4pt}{4pt}
\titlespacing*{\subsection}{0pt}{15pt}{1pt}

\fancyfoot{}
\fancyfoot[LO,RE]{\vspace{-7.1pt}\includegraphics[height=9pt]{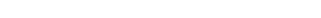}}
\fancyfoot[CO]{\vspace{-7.1pt}\hspace{13.2cm}\includegraphics{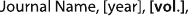}}
\fancyfoot[CE]{\vspace{-7.2pt}\hspace{-14.2cm}\includegraphics{head_foot/RF}}
\fancyfoot[RO]{\footnotesize{\sffamily{1--\pageref{LastPage} ~\textbar  \hspace{2pt}\thepage}}}
\fancyfoot[LE]{\footnotesize{\sffamily{\thepage~\textbar\hspace{3.45cm} 1--\pageref{LastPage}}}}
\fancyhead{}
\renewcommand{\headrulewidth}{0pt} 
\renewcommand{\footrulewidth}{0pt}
\setlength{\arrayrulewidth}{1pt}
\setlength{\columnsep}{6.5mm}
\setlength\bibsep{1pt}

\makeatletter 
\newlength{\figrulesep} 
\setlength{\figrulesep}{0.5\textfloatsep} 

\newcommand{\topfigrule}{\vspace*{-1pt}%
\noindent{\color{cream}\rule[-\figrulesep]{\columnwidth}{1.5pt}} }

\newcommand{\botfigrule}{\vspace*{-2pt}%
\noindent{\color{cream}\rule[\figrulesep]{\columnwidth}{1.5pt}} }

\newcommand{\dblfigrule}{\vspace*{-1pt}%
\noindent{\color{cream}\rule[-\figrulesep]{\textwidth}{1.5pt}} }

\makeatother

\twocolumn[
  \begin{@twocolumnfalse}
{\includegraphics[height=30pt]{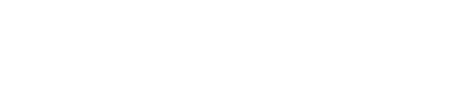}\hfill\raisebox{0pt}[0pt][0pt]{\includegraphics[height=55pt]{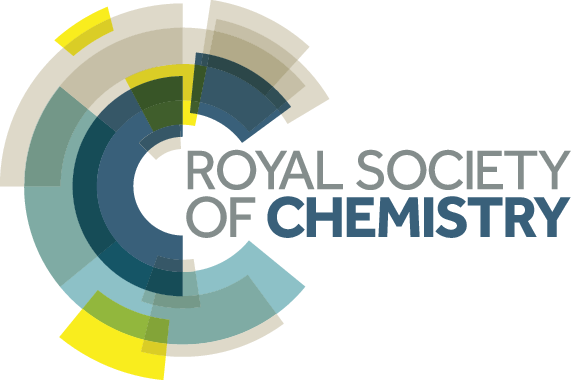}}\\[1ex]
\includegraphics[width=18.5cm]{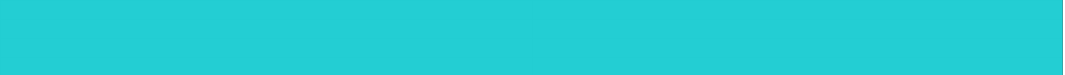}}\par
\vspace{1em}
\sffamily
\begin{tabular}{m{4.5cm} p{13.5cm} }

\includegraphics{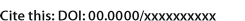} & \noindent\LARGE{\textbf{
Library of carbon nanotube junctions: data-driven insights into structure–magnetotransport relationships
}}\\
\vspace{0.01cm} & \vspace{0.01cm} \\

& \noindent\large{
Juan Alberto Canch\'e-Mart\'in,$^{\dag}$\textit{$^{1,2}$} 
Teresa Kulka,$^{\dag}$\textit{$^{3}$} 
Jacek A. Majewski,\textit{$^{3,4}$} 
Irina V. Lebedeva,\textit{$^{1,5,6}$} 
Maciej Marchwiany,$^{\ast}$\textit{$^{7}$} 
and 
Karolina Z. Milowska$^{\ast}$\textit{$^{1,8,9}$}} \\

\includegraphics{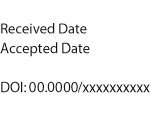} & \noindent\normalsize{Electronic transport in carbon nanotube (CNT) assemblies is controlled by a heterogeneous population of CNT--CNT junctions, yet most microscopic studies consider only a few representative systems. Here, we construct a library of 146 single-walled carbon nanotube (SWCNT)--SWCNT junctions spanning broad structural and electronic diversity and analyse their magnetotransport using an automated workflow combining molecular dynamics, tight-binding theory, Peierls magnetic coupling, and non-equilibrium Green's functions.The resulting transport data are subsequently analysed using machine-learning methods. Two complementary transport descriptors reveal distinct structure--transport hierarchies. The averaged first transmission-step value is governed primarily by the mean chiral angle of the two nanotubes forming a junction, whereas the energy gap of the junction depends predominantly on the metallic or semiconducting character of the constituent CNTs. Temperature generally suppresses  the averaged transmission while reducing the extracted gap, whereas a perpendicular magnetic field affects transmission much more strongly than the gap. Signatures of interference-driven transport remain visible even at $300~\mathrm{K}$. Notably, semiconducting--semiconducting junctions retain the largest gaps but, once shifted into their conducting regime, can exhibit transmission comparable to or exceeding that of metallic junctions. These results establish statistically robust junction-level trends that can inform future network-scale models of CNT assemblies. } \\

\end{tabular}

 \end{@twocolumnfalse} \vspace{0.6cm}
]

\renewcommand*\rmdefault{bch}\normalfont\upshape
\rmfamily
\section*{}
\vspace{-1cm}


\footnotetext{\textit{$^{1}$~CIC nanoGUNE, Tolosa Hiribidea 76, Donostia-San Sebastian, 20018, Spain}}
\footnotetext{\textit{$^{2}$~Autonomous University of Yucatan, Mexico}}
\footnotetext{\textit{$^{3}$~Faculty of Physics, University of Warsaw, Ludwika Pasteura 5, 02-093 Warsaw, Poland}}
\footnotetext{\textit{$^{4}$~Center for Terahertz Research and Applications - Centera 2, Centre for Advanced Materials and Technologies (CEZAMAT), Warsaw University of Technology, Warsaw, Poland}}
\footnotetext{\textit{$^{5}$~Simune Atomistics, Donostia-San Sebastián, 20018, Spain}}
\footnotetext{\textit{$^{6}$~Catalan Institute of Nanoscience and Nanotechnology (ICN2), CSIC and BIST, Campus UAB, Bellaterra, 08193 Barcelona, Spain}}
\footnotetext{\textit{$^{7}$~Interdisciplinary Centre for Mathematical and Computational Modelling (ICM), University of Warsaw, Pawińskiego 5a, 02-106 Warsaw, Poland, E-mail: m.marchwiany@icm.edu.pl}}
\footnotetext{\textit{$^{8}$~Ikerbasque, Basque Foundation for Science, Plaza Euskadi, Bilbao, 48009, Spain}}
\footnotetext{\textit{$^{9}$~BCMaterials, Basque Center for Materials, Applications and Nanostructures, UPV/EHU Science Park, 48940 Leioa, Spain; E-mail: kz.milowska@bcmaterials.net}}


\footnotetext{\dag~These authors contributed equally to this work}


\section{Introduction}

Macroscopic conductors assembled from low-dimensional building blocks, including nanotubes, nanowires and nanosheets, are fundamentally different from homogeneous bulk metals. Their electrical response emerges from transport through networks of individual nano-objects connected by a large number of structurally distinct junctions. Even when network connectivity is well established, transport can remain strongly inhomogeneous, and the resistance associated with interparticle junctions can substantially limit the macroscopic conductivity~\cite{Gabbett2024}. At the network scale, such contacts are often represented by effective junction resistances~\cite{Topinka2009,Gabbett2024}. Microscopically, however, a junction is not a passive resistor with a fixed resistance: its electronic transmission depends on the electronic structures of the constituent materials, their local atomic arrangement, and external conditions. Quantum interference can therefore make the response of an individual junction strongly energy-, geometry-, temperature-, and field-dependent~\cite{Taming}.

Carbon nanotube (CNT) assemblies~\cite{Bulmer2017,Bulmer,manuscriptAGA,Taming} provide a particularly useful system in which to investigate this problem. Individual CNTs have well-defined atomic structures specified by their chirality, which determines their diameter and electronic character, while macroscopic CNT films and fibres naturally consist of large networks of CNTs connected through numerous intertube contacts. Their electrical properties therefore combine well-defined nanoscale electronic building blocks with the structural heterogeneity characteristic of realistic nanomaterial networks~\cite{Xiang2025}. This makes CNT assemblies an attractive model for determining how microscopic junction structure translates into broader transport trends.

A substantial body of experimental and theoretical work has therefore focused on transport through individual CNT--CNT contacts. Experiments have demonstrated pronounced dependences of junction conductance on crossing angle and overlap~\cite{Barnett2019}, while controlled parallel contacts show a strong variation of resistance with contact length~\cite{Hamasaki2022}. Atomistic calculations have further identified important roles of nanotube chirality, atomic registry, intertube separation, structural relaxation, and crossing geometry \cite{Xu2013,Tripathy2016,Durrant2022,Adinehloo2023}. The latter can modify the electronic junction conductance by orders of magnitude, emphasising that nominally similar CNT contacts can represent very different electronic transport elements. Despite this progress, a systematic picture of how the different structural and electronic characteristics of CNT junctions compete in determining transport is still missing. Existing microscopic studies have typically explored only restricted regions of this large parameter space. Extensive sampling of junction geometry has revealed clear roles of contact area and interatomic separation, but has largely been performed for a single metallic armchair chirality~\cite{Durrant2022}. Conversely, studies spanning different CNT types have established strong effects of chirality, registry, crossing angle, overlap, and intertube coupling, but generally for a limited number of selected junctions~\cite{Adinehloo2023,Wittemeier2022}. As a result, it remains unclear which of these descriptors dominate when structural and electronic diversity are considered simultaneously across a broad population of CNT--CNT junctions.

Most junction-resolved studies have also focused primarily on zero-field transport. A magnetic field introduces an additional degree of control because its orbital effect modifies the phases accumulated along electronic propagation paths and can therefore change the interference conditions within a junction. In a perpendicular field, this effect acts directly on the intertube transport region and can produce strongly junction-specific changes in transmission~\cite{Taming}. Magnetotransport therefore provides access to aspects of junction physics that are not captured by zero-field conductance alone.
Our recent studies examined this connection between microscopic junction physics and magnetotransport using selected representative systems. In Ref.~\cite{manuscriptAGA}, an atomistic framework combining quantum transport, finite-temperature structural fluctuations, and magnetic-field effects was used to identify junction-level mechanisms underlying the magnetotransport of CNT assemblies. In particular, the calculations revealed distinct roles of junction overlap and lattice mismatch in determining the magnetic response. More recently, Ref.~\cite{Taming} resolved the coherent transport mechanism in individual CNT junctions in greater detail. CNT--CNT contacts were shown to behave as coupled electronic waveguides, in which intertube coupling and quantum interference produce strongly energy-, geometry-, and magnetic-field-dependent transmission. 
These studies established the microscopic mechanisms governing selected CNT junctions, but they leave open a different question: which of these structure--transport relationships remain statistically important when a much broader population of junctions is considered? Real SWCNT assemblies contain distributions of diameters, chiralities, and electronic characters, so conclusions drawn from a small number of carefully selected junctions cannot necessarily be extrapolated across the full junction population. It is also not known to what extent the interference-driven magnetic response identified for selected junctions remains detectable across such structural diversity and at finite, including room, temperature.

Here, we address these questions using a structurally and electronically diverse library of 146 SWCNT--SWCNT junctions constructed from 85 distinct nanotube chiralities. For each junction, transport was evaluated at six temperatures using 12 independent MD-generated configurations per temperature and 50 perpendicular magnetic-field values. The resulting dataset comprises 10,512 finite-temperature structural realisations and more than half a million magnetic-field-resolved transport calculations. An automated atomistic transport workflow was used to generate the dataset, which was subsequently analysed using machine-learning methods to identify relationships between junction structure and transport. We focus on two complementary transport characteristics: the averaged first transmission-step value and the energy gap. We show that they are governed by distinctly different hierarchies of junction descriptors. The efficiency of transmission once propagating states become available is associated most strongly with the mean chiral angle of the nanotube pair, whereas the junction energy gap depends predominantly on the metallic or semiconducting character of the constituent CNTs. The analysis also demonstrates that the interference-driven magnetic-field response previously identified in selected junctions persists across a much broader structural space and remains detectable even at room temperature.

\section{Methods}

\subsection{Construction of the library SWCNT--SWCNT junctions }

Experimental SWCNT assemblies prepared from unsorted material commonly contain mixtures of nanotubes with different diameters, chiralities and electronic characters \cite{Yang2023}. Although chirality-selective synthesis and post-synthesis separation provide access to enriched and single-chirality samples, obtaining large quantities of a prescribed chirality at high purity remains experimentally demanding \cite{Dzienia2024,Yomogida2016,Yang2023}. We therefore constructed a junction library spanning this structural and electronic diversity, rather than reproducing the composition of a particular experimental sample.

Candidate SWCNTs were restricted to diameters between 0.5 and 3.0\,nm, providing a broad sampling window that includes experimentally accessible subnanometre nanotubes and larger-diameter species \cite{Dzienia2024,Hwang2026}. Nanotubes containing more than 500 atoms per translational unit cell were excluded to limit the computational cost of the subsequent junction calculations. 
For the subsequent analysis, SWCNTs were classified as metallic (M) when their chiral indices satisfied $(n-m)=3k$, with integer $k$, and as semiconducting (SC) otherwise. Here, the metallic designation follows the zone-folding criterion and does not exclude small curvature-induced gaps in non-armchair nanotubes. During library construction, armchair and non-armchair metallic nanotubes were sampled separately to ensure representation of both groups, but were combined into a single metallic class for the analysis. Farthest-point sampling based on nanotube diameter and chiral angle was used to avoid clustering of the selected nanotubes in parameter space.

SWCNT pairs were subsequently selected to construct a balanced and structurally diverse library of SWCNT--SWCNT junctions.  Each nanotube was first assigned to one of three diameter classes: small ($2R<1~\mathrm{nm}$), medium ($1~\mathrm{nm}\leq2R<2~\mathrm{nm}$), or big ($2R\geq2~\mathrm{nm}$), and to one of three chiral-angle classes: $\theta=0^{\circ}$, $0^{\circ}<\theta<30^{\circ}$, or $\theta=30^{\circ}$. Target numbers of junctions for the metallic--metallic, metallic--semiconducting, and semiconducting--semiconducting  combinations were then specified, while enforcing a minimum number of metallic--metallic junctions and reserving a number of self-junctions, formed by two SWCNTs of identical chirality, since these would otherwise be poorly represented.
The specific SWCNT pairs were selected using a stratified greedy procedure. For each required electronic combination, the algorithm searched the available candidate pairs and preferentially selected those whose combinations of the diameter classes and, separately, the chiral-angle classes defined above were underrepresented in the selected set. In this way, the selected junctions were distributed across the corresponding $3\times3$ diameter and chiral-angle grids rather than clustering in a limited number of combinations. 
After the required cross-chirality pairs had been selected, the reserved self-junctions were added at random.

Following the initial automated selection, additional junctions were introduced to improve the representation of under-sampled regions of the library, in particular junctions containing medium-diameter nanotubes and metallic--metallic combinations. 
The final library contains 146 SWCNT--SWCNT junctions constructed from 85 distinct SWCNT chiralities: 44 metallic and 41 semiconducting nanotube types. The junction set comprises 60 metallic--metallic, 47 metallc--semiconducting, and 39 semiconducting--semiconducting junctions. All SWCNT types included in the library are shown in Fig.~\ref{fig:CNTs} and The library-construction procedure is summarised schematically in the first stage of Fig.~\ref{fig:workflow}.

\subsection{
TB--NEGF--MD--Peierls framework
}

\begin{figure}[h!tb]
    \centering
    \includegraphics[width=0.82\linewidth]{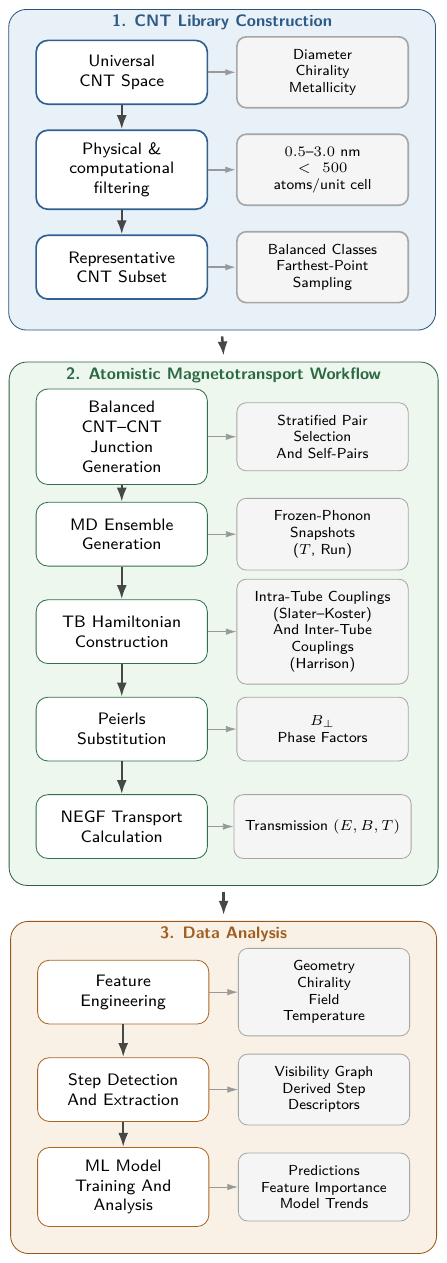}
    \caption{
Workflow for constructing the CNT--CNT junction library, computing the electronic transport response of the junctions under external magnetic fields and at finite temperatures, and analysing the resulting transport descriptors using ML-based methods.
    }
    \label{fig:workflow}
\end{figure}

Electronic transport through the SWCNT--SWCNT junction library, including
the effects of finite temperature and an external magnetic field, was studied using the TB--NEGF--MD--Peierls framework introduced in our previous work \cite{manuscriptAGA}. The framework combines a tight-binding (TB) description of the electronic structure with non-equilibrium Green's function (NEGF) transport calculations, molecular dynamics (MD) sampling of finite-temperature atomic configurations, and Peierls substitution to account for the orbital effect of an external magnetic field.

In the present work, the TB--NEGF--MD--Peierls framework was incorporated into an automated high-throughput procedure that constructs the prescribed SWCNT--SWCNT junction devices and performs the corresponding transport calculations over predefined sets of temperatures and perpendicular magnetic fields. The overall procedure is summarised in the second stage of Fig.~\ref{fig:workflow}. 
For each junction and temperature, independent MD simulations were performed in the QuantumATK numerical package \cite{Smidstrup2020} to generate thermally perturbed atomic
configurations, which were subsequently treated as frozen-phonon snapshots. The subsequent electronic transport calculations were performed using our modified implementations of \texttt{sisl}, based on v0.15.1 \cite{PapiorSisl2023}, and \texttt{TBtrans}, based on \texttt{SIESTA} 5.2.0-alpha \cite{Papior2017}. The modifications, described in Ref.~\cite{Taming}, enable the treatment of complex-valued Hamiltonian and overlap matrices in both the central scattering region and the semi-infinite electrodes. This allows the magnetic-field-induced Peierls phase to be included consistently in both matrices throughout the junction device and its electrodes. For each snapshot, the non-orthogonal TB Hamiltonian and overlap matrices were constructed using the modified \texttt{sisl} implementation. The orbital effect of the perpendicular magnetic field was introduced through the Peierls phase, and the corresponding energy-resolved transmission spectra were calculated within the NEGF formalism using the modified \texttt{TBtrans} implementation. The transmission spectra obtained for the thermal realisations were then ensemble averaged and processed to extract the electronic transport descriptors used in the subsequent statistical and ML-based analysis.

\subsubsection{Tight-binding calculations}


\paragraph{
Intra-nanotube Hamiltonian}


The electronic structure within each SWCNT was described using an empirical TB model with one $p$ orbital per carbon atom, corresponding to the $p_z$ orbital of planar graphene. These orbitals form the $\pi$ and $\pi^*$ bands that dominate the low-energy electronic structure near the Fermi level~\cite{Dresselhaus1996}. 
The TB Hamiltonian was constructed following the Slater--Koster approximation, which expresses the coupling matrix elements between atomic orbitals in terms of transfer integrals and orbital orientations. In the case of CNTs, the curvature of the cylindrical surface breaks the planar symmetry present in graphene, thereby modifying the relative orientations and coupling of the $p_z$ orbitals. Curvature therefore introduces a $\sigma$-type contribution to hopping between neighbouring $p_z$ orbitals, in addition to the $\pi$-type contribution \cite{Klinovaja2011}.
%
To capture this effect, the geometric dependence of the hopping integrals was included through the relative orbital orientations and their projections onto the interatomic bond \cite{SlaterKoster1954}.
Formally, the intra-nanotube TB Hamiltonian can be written as
%
%
\begin{equation}
\begin{aligned}
H_{\mathrm{intra}} ={}&\sum_{\mu}\epsilon_{\mu}\, c_{\mu}^{\dagger}c_{\mu} +\sum_{\langle\mu,\nu\rangle} t_{\mu\nu}\left( c_{\mu}^{\dagger}c_{\nu} +\mathrm{h.c.} \right), 
\end{aligned}
\label{eq:H_intra}
\end{equation}
where $\epsilon_{\mu}$ is the on-site energy of orbital $\mu$, and $t_{\mu\nu}$ is the hopping integral between atoms $\mu$ and $\nu$. The operators $c_{\mu}^{\dagger}$ and $c_{\mu}$ create and annihilate an electron in orbital $\mu$, respectively. The pair sum includes the interacting atoms within each nanotube, with each pair counted once, and $\mathrm{h.c.}$ denotes the Hermitian conjugate of the hopping term. Only nearest-neighbour intra-nanotube hopping and orbital overlap were retained, using an interatomic-distance cutoff of $1.52$~\AA. 
%
%
According to the Slater–Koster approximation, the hopping between two $p$ orbitals depends on their orientation relative to the interatomic bond:
\begin{equation}
\begin{aligned}
t_{\mu\nu} = (\mathbf{n}_\mu \cdot \mathbf{n}_\nu) V_{pp\pi} + \frac{(\mathbf{n}_\mu \cdot \mathbf{R}_{\nu\mu})(\mathbf{n}_\nu \cdot \mathbf{R}_{\nu\mu})}{R_{\nu\mu}^2} (V_{pp\sigma} - V_{pp\pi})
\end{aligned}
\label{eq:SK_intra}
\end{equation}
where $\mathbf{n}_{\mu}$ and $\mathbf{n}_{\nu}$ are unit vectors along the local $p_z$ orbital directions at atoms $\mu$ and $\nu$, $\mathbf{R}_{\nu\mu}=\mathbf{r}_{\nu}-\mathbf{r}_{\mu}$ is the vector connecting the two atoms, $R_{\nu\mu}=|\mathbf{R}_{\nu\mu}|$, and $V_{pp\pi}$ and $V_{pp\sigma}$ are the Slater--Koster transfer integrals.
In the nanotube geometry, the local $p_z$ orbital directions were approximated by radial unit vectors perpendicular to the $z$ axis. For each atomic configuration, the reference axis of each nanotube was taken parallel to $z$ through its mean transverse atomic position. These radial vectors coincide with the surface normals for an ideal circular nanotube, but do not represent reconstructed local surface normals for a distorted nanotube. 
The hopping depends on both the relative alignment of the orbital directions and their projections onto the interatomic bond.
The curvature contribution is therefore contained in the scalar products $\mathbf{n}_{\mu}\cdot\mathbf{n}_{\nu}$, $\mathbf{n}_{\mu}\cdot\mathbf{R}_{\nu\mu}$, and $\mathbf{n}_{\nu}\cdot\mathbf{R}_{\nu\mu}$ appearing in Eq.~\ref{eq:SK_intra}.  Accounting for orbital orientation is particularly relevant for small-diameter nanotubes \cite{Klinovaja2011} and structurally deformed junctions. In the flat graphene limit, the orbital directions are parallel and perpendicular to the interatomic bond, so the hopping expression reduces to $t_{\mu\nu}=V_{pp\pi}$.
%
%
%
The non-orthogonality of the basis was represented by the overlap matrix $\mathbf{S}$. Its off-diagonal intra-nanotube elements were evaluated using the same angular expression as the hopping integrals in Eq.~\ref{eq:SK_intra}, with $V_{pp\pi}$ and $V_{pp\sigma}$ replaced by $s_{pp\pi}$ and $s_{pp\sigma}$, respectively. The hopping parameters ($V_{pp\pi}=-2.381~\mathrm{eV}$ and
$V_{pp\sigma}=8.389~\mathrm{eV}$) follow the Harrison tight-binding
parametrisation~\cite{Harrison1981,Harrison1999}, with the
specific numerical values used here communicated in Ref.~\cite{AndreaSF}. The overlap parameters ($s_{pp\pi}=0.129$ and $s_{pp\sigma}=0.146$) were taken from Ref.~\cite{Treue2010}.  These parameters were held constant, with no additional bond-length scaling applied to intra-nanotube interactions. Within the retained neighbour set, the geometrical dependence of the hopping and overlap matrix elements therefore arises from the angular factors. 


\paragraph{
Inter-nanotube coupling}

To describe the interaction between CNTs in a junction, a complementary approach based on a Harrison-type hopping parametrisation was employed. This parametrisation introduces an interatomic-distance dependence into the transfer integrals between atoms belonging to different CNTs, while retaining the Slater--Koster dependence on orbital orientation.
%
In the present model, the inter-nanotube transfer integrals were assumed to follow an inverse-square distance dependence:
\begin{equation}
V_{pp\lambda}^{\mathrm{inter}}(d_{\mu\nu}) = V_{pp\lambda} \left(\frac{a_{\mathrm{C}}}{2d_{\mu\nu}}\right)^2,
\label{eq:inter_hopping_scaling}
\end{equation}
where $\lambda=\pi,\sigma$, $a_{\mathrm{C}}=1.42$~\AA\ is the reference carbon--carbon bond length, and $d_{\mu\nu}=|\mathbf{R}_{\nu\mu}|$ is the distance between atoms $\mu$ and $\nu$ belonging to different CNTs. 
The inter-nanotube hopping matrix elements $t_{\mu\nu}^{\mathrm{inter}}$ were evaluated using the angular expression in Eq.~\ref{eq:SK_intra}, with $V_{pp\pi}$ and $V_{pp\sigma}$ replaced by their distance-dependent inter-nanotube values. The orbital directions were evaluated using the radial approximation described above, with each atom referenced to the axis of its respective nanotube. Thus, the inter-nanotube hopping depends on both the interatomic distance and the relative orbital orientations.
%
The inter-nanotube overlap parameters were scaled according to Eq.~\ref{eq:inter_hopping_scaling}, with $V_{pp\lambda}$ replaced by $s_{pp\lambda}$ for $\lambda=\pi,\sigma$. The corresponding overlap matrix elements were obtained using the same angular expression as the hopping matrix elements, with the hopping parameters replaced by the overlap parameters.
Inter-nanotube hopping and overlap were retained up to an interatomic-distance cut-off of $3.60$~\AA. Contributions beyond this distance were neglected. Within the retained interaction range, the Harrison-type scaling accounts for the decrease in the magnitudes of the transfer integrals with increasing interatomic distance.


This procedure is especially relevant for junction studies, since transport properties depend not only on the electronic structure of each individual nanotube but also on the coupling terms between them. 
The inter-nanotube contribution to the Hamiltonian is
\begin{equation}
H_{\mathrm{inter}} = \sum_{\mu\in\mathrm{CNT}_1} \sum_{\nu\in\mathrm{CNT}_2} \left( t_{\mu\nu}^{\mathrm{inter}}\, c_{\mu}^{\dagger}c_{\nu} +\mathrm{h.c.} \right), 
\label{eq:H_inter}
\end{equation}
where $\mu$ and $\nu$ label atoms belonging to the first and second nanotubes, respectively. The hopping matrix elements $t_{\mu\nu}^{\mathrm{inter}}$ were obtained from Eq.~\ref{eq:SK_intra} using the distance-dependent transfer integrals defined in Eq.~\ref{eq:inter_hopping_scaling}, and were set to zero beyond the inter-nanotube cut-off. Each inter-nanotube pair is counted once, with the Hermitian conjugate accounting for the reverse hopping.

Thus, the total Hamiltonian of the junction is composed of the two main contributions:
\begin{equation}
H_{\mathrm{junction}} = H_{\mathrm{intra}}+H_{\mathrm{inter}},
\label{eq:H_junction}
\end{equation}
where $H_{\mathrm{intra}}$ contains the on-site and intra-nanotube hopping terms of both constituent CNTs, and $H_{\mathrm{inter}}$ describes the hopping between them.
%
The combination of these contributions incorporates the effects of orbital orientation, relative atomic arrangement, and interatomic separation on electronic coupling within and between the CNTs.

\subsubsection{Magnetotransport calculations}

\begin{figure*}
     \centering
     \includegraphics[width=1.0\linewidth]{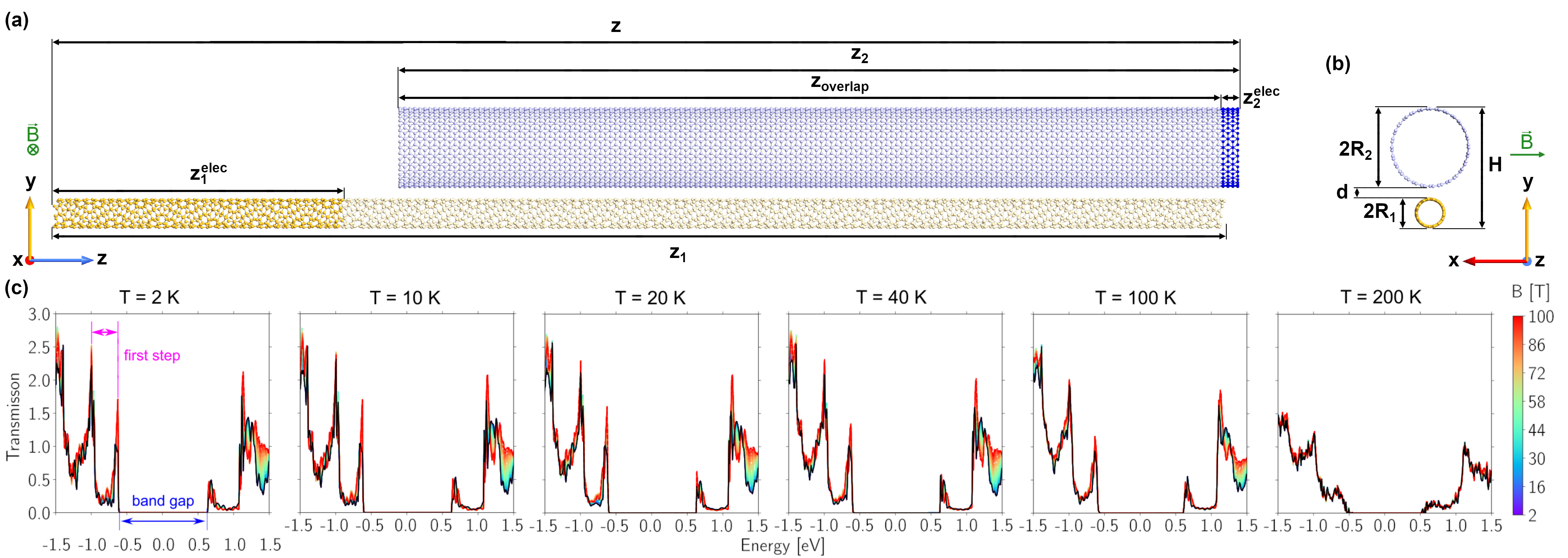}
     \caption{Representative CNT-CNT junction. (a) Side and (b) cross-sectional views of the $(8,1)+(14,14)$ junction used in the magnetotransport calculations. The semiconducting $(8,1)$ nanotube (yellow) is connected to a left semi-infinite electrode of the same chirality (dark yellow), whereas the metallic $(14,14)$ nanotube (blue) is connected to a right semi-infinite electrode of the same chirality (navy blue). The left and right nanotubes are described by the chiral vectors $(\mathrm{n}_{11},\mathrm{n}_{12})$ and $(\mathrm{n}_{21},\mathrm{n}_{22})$ and chiral angles $\theta_{1}$ and $\theta_{2}$, respectively; for the system shown, $\mathrm{n}_{11}=8$, $\mathrm{n}_{12}=1$, $\mathrm{n}_{21}=\mathrm{n}_{22}=14$, $\theta_{1}=5.82^{\circ}$, and $\theta_{2}=30^{\circ}$. The nanotube radii R$_{1}$ and R$_{2}$, electrode lengths z$_{1}^{\mathrm{elec}}$ and z$_{2}^{\mathrm{elec}}$, junction overlap length z$_{\mathrm{overlap}}$, and intertube distance d are indicated. (c) Energy-resolved ensemble-averaged transmission spectra of the $(8,1)+(14,14)$ CNT junction with an overlap length of $z_{\mathrm{overlap}}=22.12~\mathrm{nm}$. Each averaged spectrum was obtained from 12 transmission spectra calculated for independent MD-generated thermal realisations at the same temperature and magnetic-field value. Black lines show the zero-field averaged transmission, whereas coloured lines correspond to perpendicular magnetic fields up to $100~\mathrm{T}$. The first transmission step on the p-doped side of the spectrum, below the Fermi level, and band gap used in the subsequent analysis are indicated for the representative averaed spectrum. 
     }
     \label{fig:system}
\end{figure*}

The orbital effect of the external magnetic field was included in both the Hamiltonian and overlap matrices, $\mathbf{H}$ and $\mathbf{S}$. It was taken into account through the Peierls substitution~\cite{Peierls1933,SaitoDresselhaus1998,PhysRevB.103.045402,Goldman2016}, which introduces a phase factor into the hopping parameters $t_{\mu\nu}$ and overlap matrix elements $s_{\mu\nu}$:
%
\begin{equation}
    t_{\mu\nu}
    \;\rightarrow\; t_{\mu\nu}e^{i\varphi},
    \label{eq:peierlsH}
\end{equation}
\begin{equation}
    s_{\mu\nu}
    \;\rightarrow\; s_{\mu\nu}e^{i\varphi},
    \label{eq:peierlsS}
\end{equation}
where
\begin{equation}
    \varphi=\frac{2\pi}{\Phi_0}\int_{\mathbf r_\mu}^{\mathbf r_\nu} \mathbf A\cdot d\mathbf r.
    \label{eq:peierlsphase}
\end{equation}
Here, $\mathbf{r}_{\mu}=(x_{\mu},y_{\mu},z_{\mu})$ denotes the position of atom $\mu$, $\Phi_0=h/e$ is the magnetic flux quantum, with $h$ being Planck's constant and $e$ the magnitude of the electron charge, and $\mathbf{A}$ is the magnetic vector potential. The full exponential phase factor was retained without expansion in the magnetic field.


The CNT axes were oriented along the $z$ direction, while the intertube displacement was taken along $y$, as shown in Fig.~\ref{fig:system}. A homogeneous magnetic field $\mathbf{B}=(B_x,0,0)$ was applied along the $x$ direction, perpendicular to both the nanotube axes and the effective junction plane. We chose the gauge $\mathbf{A}=(0,0,yB_x)$, which preserves translational invariance along the $z$ direction. The corresponding Peierls phase is 
\begin{equation}
    \varphi=\frac{\pi}{\Phi_0}B_x(z_\nu - z_\mu)(y_\mu + y_\nu).
    \label{eq:phase-f}
\end{equation}
The perpendicular magnetic-field orientation used here corresponds to the maximum-flux configuration considered in Ref.~\cite{Taming}, maximising the orbital response for a given junction geometry. The Peierls phase was evaluated using the actual atomic coordinates of each thermally perturbed configuration.

The calculations were spin-unpolarised and the Zeeman term was neglected. Its effect on the transmission was analysed explicitly in Ref.~\cite{Taming}, where it was shown that, in the absence of spin--orbit coupling, it produces two independent spin contributions shifted in energy. Here, we focus on the orbital magnetic-field response introduced through the Peierls phase.

In that way, we create magnetic-field-dependent TB Hamiltonian and overlap matrices that describe most important effects of the junction geometry. Our model provides a solid foundation for the subsequent calculation of the electronic transport properties using non-equilibrium Green’s function (NEGF) formalism \cite{Papior2017, PhysRevB.65.165401}, ensuring consistency with the underlying atomic-scale physics and with a level of simplification suitable for tackling large-scale systems.

The resulting magnetic-field-dependent TB Hamiltonian and overlap matrices were used to calculate the electronic transport properties within the non-equilibrium Green's function (NEGF) formalism \cite{Papior2017,PhysRevB.65.165401}. The junction device was divided into a semi-infinite left electrode $L$, a central scattering region $C$, and a semi-infinite right electrode $R$. The energy-resolved transmission from the left to the right electrode was calculated using the Landauer--B\"uttiker expression
\begin{equation}
\begin{aligned}
T_{L\rightarrow R}(E,B) = \mathrm{Tr}\big[ \mathbf{\Gamma}_{R}(E,B)\mathbf{G}_{C}(E,B)  &\times \mathbf{\Gamma}_{L}(E,B) \mathbf{G}_{C}^{\dagger}(E,B) \big],
\end{aligned}
\label{eq:transmission}
\end{equation}
where $\mathbf{\Gamma}_{\alpha}$, with $\alpha=L,R$, is the electrode broadening matrix
\begin{equation}
\mathbf{\Gamma}_{\alpha}(E,B) = i\left[ \mathbf{\Sigma}_{\alpha}(E,B) - \mathbf{\Sigma}_{\alpha}^{\dagger}(E,B) \right],
\label{eq:scattering}
\end{equation}
and $\mathbf{G}_{C}$ is the retarded Green's function of the central region coupled to the electrodes. 
For brevity, the magnetic-field dependence of the matrices is omitted in the following expressions. The Green's function is given by
\begin{equation}
    \mathbf G_C^{-1}(E)= (E+i\eta) \mathbf S_C - \mathbf H_C - \mathbf \Sigma_L(E)-\mathbf \Sigma_R(E),
    \label{eq:G}
\end{equation}
where $\mathbf{H}_{C}$ and $\mathbf{S}_{C}$ are the Hamiltonian and overlap matrices of the central region, respectively, and $\eta\rightarrow0^{+}$ defines the retarded Green's function.
The self-energies account for the effect of the semi-infinite electrodes on the finite scattering region and arise from integrating out the electrode degrees of freedom. The self-energy of electrode $\alpha=L,R$ is
\begin{equation}
    \mathbf \Sigma_{\alpha}(E)=(\mathbf H_{C\alpha}-E \mathbf S_{C\alpha}) \mathbf G_{\alpha}(E)(\mathbf H_{C\alpha}^{\dagger}-E \mathbf S_{C\alpha}^{\dagger}),
    \label{eq:sigma}
\end{equation}
where $\mathbf{G}_{\alpha}$ is the surface Green's function of the corresponding semi-infinite electrode. Formally,
\begin{equation}
    \mathbf G_{\alpha}^{-1}(E)= (E+i\eta) \mathbf S_{\alpha} - \mathbf H_{\alpha},
    \label{eq:Glead}
\end{equation}
with the surface Green's function evaluated using the Sancho--Rubio iterative method.
For the coupling between the central region and a given electrode $\alpha$, the Hamiltonian and overlap matrices can be written in block form as
\begin{equation}
\mathbf{H} =
\begin{pmatrix}
\mathbf{H}_{C} & \mathbf{H}_{C\alpha} \\
\mathbf{H}_{\alpha C} & \mathbf{H}_{\alpha}
\end{pmatrix},
\qquad
\mathbf{S} =
\begin{pmatrix}
\mathbf{S}_{C} & \mathbf{S}_{C\alpha} \\
\mathbf{S}_{\alpha C} & \mathbf{S}_{\alpha}
\end{pmatrix},
\qquad
\alpha=L,R.
\label{eq:HSblocks}
\end{equation}
Here, $\mathbf{H}_{\alpha}$ and $\mathbf{S}_{\alpha}$ describe the corresponding semi-infinite electrode, while $\mathbf{H}_{C\alpha}$ and $\mathbf{S}_{C\alpha}$ describe its coupling to the central region. Hermiticity of the complete Hamiltonian and overlap matrices requires $\mathbf{H}_{\alpha C}=\mathbf{H}_{C\alpha}^{\dagger}$ and $\mathbf{S}_{\alpha C}=\mathbf{S}_{C\alpha}^{\dagger}$.

Relative to the zero-field matrices, the magnetic field introduces changes $\delta\mathbf{H}$ and $\delta\mathbf{S}$ through the Peierls phase. The electrode self-energies consequently also become magnetic-field dependent. The central-region Green's function can therefore be written as
\begin{equation}
\begin{aligned}
\mathbf{G}_{C}^{-1}(E,B) ={} (E+i\eta) \left[ \mathbf{S}_{C} + \delta\mathbf{S}_{C}(B) \right]
- \left[ \mathbf{H}_{C} + \delta\mathbf{H}_{C}(B) \right]  \\- \left[ \mathbf{\Sigma}_{L}(E) + \delta\mathbf{\Sigma}_{L}(E,B) \right] - \left[ \mathbf{\Sigma}_{R}(E) + \delta\mathbf{\Sigma}_{R}(E,B) \right].
\end{aligned}
\label{eq:dG}
\end{equation}

In the present calculations, the Peierls phase was included in the Hamiltonian and overlap matrices of both electrodes and the central scattering region, as well as in the corresponding coupling blocks. The magnetic-field-dependent matrices were constructed using \texttt{sisl}, and the transmission spectra were calculated using \texttt{TBtrans}. The underlying NEGF--Peierls implementation and its application to CNT junctions were described and benchmarked previously in Ref.~\cite{Taming}.

\subsubsection{Finite-temperature transport calculations}

To include finite-temperature structural fluctuations, we employed the MD--Landauer approach, which combines classical molecular dynamics (MD) sampling with electronic transport calculations within the Landauer formalism~\cite{manuscriptAGA}. The MD simulations generate thermally perturbed atomic configurations that include anharmonic structural fluctuations within the classical interatomic potential. Each configuration was subsequently treated as a frozen-phonon snapshot, so that electron transport through a given atomic configuration remained elastic.

For each junction and temperature, 12 separate MD realisations were generated. Each realisation was obtained in two stages. First, NVT Langevin MD simulations were performed separately for the left and right electrodes, with their centres of mass fixed. The final electrode configurations were then used to construct the complete junction device. In the second stage, an NVT Langevin MD simulation was performed for the full device. During this simulation, the left-electrode atoms were fixed, while the right electrode was constrained to move as a rigid body. The final configuration from each full-device MD run was used for the subsequent transport calculations.
%
The MD simulations were performed using the Langevin thermostat \cite{Collins2017} and the Tersoff-C-1989 interatomic potential \cite{Tersoff1988}, as implemented in the QuantumATK numerical package \cite{Schneider2017,Smidstrup2020,QuantumATK2022}. Initial atomic velocities were assigned according to the Maxwell--Boltzmann distribution. Both MD stages were performed using a time step of $0.1~\mathrm{fs}$ for $50\,000$ steps, corresponding to a simulation time of $5~\mathrm{ps}$ per stage. The default Langevin friction constant of $0.01~\mathrm{fs}^{-1}$ was used. Calculations were performed at $2$, $10$, $20$, $40$, and $100~\mathrm{K}$ for all junctions. For most junctions, the highest temperature considered was $300~\mathrm{K}$, while for selected systems the highest-temperature calculation was instead performed at $200~\mathrm{K}$. Representative temperature-dependent transmission spectra are shown in Fig.~\ref{fig:system}(c) for a junction calculated up to $200~\mathrm{K}$ and in Fig.~\ref{fig:Tmetallic}(c) for a junction calculated up to $300~\mathrm{K}$.

For each thermally perturbed configuration, the corresponding TB Hamiltonian and overlap matrices were constructed and the energy-resolved transmission was calculated within the NEGF formalism for every considered magnetic-field value. The same set of 12 MD-generated configurations at a given temperature was therefore evaluated over the complete magnetic-field range. The ensemble-averaged transmission was calculated as
\begin{equation}
\overline{T}(E,B,\mathrm{T}) = \frac{1}{N_{\mathrm{MD}}} \sum_{m=1}^{N_{\mathrm{MD}}}
T_{m} \left[E,B;\mathbf{R}^{(m)}(\mathrm{T})\right], \qquad N_{\mathrm{MD}}=12,
\label{eq:Tavg}
\end{equation}
where $\mathrm{T}$ denotes temperature and $\mathbf{R}^{(m)}(\mathrm{T})$ is the atomic configuration obtained from the $m$th MD realisation at that temperature. The resulting ensemble-averaged transmission spectra $\overline{T}(E,B,\mathrm{T})$ were used directly in the subsequent analysis of the transport descriptors.
Because the MD sampling is classical, zero-point motion and quantum phonon statistics are not included. Moreover, the frozen-phonon treatment does not explicitly describe inelastic phonon-assisted transitions involving energy exchange between electrons and lattice vibrations.

\subsection{Machine learning data analysis}

\begin{table}[h!]
\centering
\begin{tabular}{l p{6cm}}
\hline
\textbf{Descriptor
} & \textbf{Description} \\
\hline
$\mathrm{z/H}$ & Ratio of the device length to its transverse height \\
F & Flatness of the device, defined as  $\mathrm{R^{min}/R^{max}}$ \\
$\mathrm{H}$ & 
Transverse height of the junction\\
$\mathrm{z}$  & Length of the device \\
$\Delta\mathrm{z}$ & Difference between the nanotube lengths  \\
$\mathrm{\Delta\mathrm{R}}$  & Difference between the nanotube radii \\
$\mathrm{z_{overlap}}$ & Overlap length \\
$\mathrm{n_{11}-n_{21}}$ & Difference between the first chiral indices
of the two nanotubes \\
$\mathrm{n_{12}-n_{22}}$ & Difference between the second chiral indices
of the two nanotubes \\
\hline
$|\Delta \uptheta|$ & Absolute difference between chiral angles of the two nanotubes \\
$\mathrm{\uptheta_{\mathrm{min}}/\uptheta_{\mathrm{max}}}$ & Ratio of the smaller to the larger chiral angle \\
$\mathrm{<\uptheta>}$  & Mean chiral angle of the two nanotubes \\
$\mathrm{n_{11}}$ & First chiral index of the left nanotube \\
$\mathrm{n_{12}}$ & Second chiral index of the left nanotube \\
$\mathrm{n_{12}/n_{11}}$ & Ratio of the second to the first chiral index
of the left nanotube \\
$<\mathrm{n_{1}}>$ & Mean value of the first chiral index for
the two nanotubes \\
$\mathrm{n_{21}}$ & First chiral index of the right nanotube \\
$\mathrm{n_{22}}$ & Second chiral index of the right nanotube \\
$\mathrm{n_{22}/n_{21}}$ & Ratio of the second to the first chiral index
of the right nanotube \\
$<\mathrm{n_{2}}>$ & Mean value of the second chiral index for
the two nanotubes \\
\hline
$\mathrm{R_{1}}$ & Radius of the left nanotube \\
$\mathrm{z_{1}}$ & Length of the left nanotube \\
F$_1$ & Flatness of the left nanotube, defined as $\mathrm{R_1^{min}/R_1^{max}}$  \\
$\mathrm{R_{2}}$ & Radius of the right nanotube \\
$\mathrm{z_{2}}$ & Length of the right nanotube \\
F$_2$  & Flatness of the right nanotube, defined as $\mathrm{R_2^{min}/R_2^{max}}$ \\
\hline
$\mathrm{M_{1}}$ & Metallicity indicator of the left nanotube; $M=1$ for metallic and $M=0$ for semiconducting \\
$\mathrm{M_{2}}$ & Metallicity indicator of the right nanotube; $M=1$ for metallic and $M=0$ for semiconducting \\
$\mathrm{M}$ &  Junction metallicity indicator; $M=1$ if at least one
nanotube is metallic and $M=0$ otherwise \\
\hline
T & Temperature \\
B & External magnetic-field strength \\
\hline
\end{tabular}
\caption{
Summary of the input descriptors used in the ML analysis.}
\label{tab:features}
\end{table}

Machine learning (ML) algorithms were used to analyse the numerical results, as schematically presented in stage 3 of Fig.~\ref{fig:workflow}. To better understand the underlying physical trends, we examined both model performance and the relative importance of the input descriptors. Based on our knowledge of the field, previous experience and the available data, we constructed the set of input descriptors (features) listed in Table~\ref{tab:features}. These included nanotube chiral indices, junction overlap length and intertube separation, together with the external conditions of temperature and magnetic field. Additional descriptors included nanotube radii, chiral angles, metallicity indicators and measures of geometrical deformation, as well as combinations of these quantities. The relevant geometrical parameters are illustrated in Fig.~\ref{fig:system}. The relative importance of the input descriptors was evaluated using random-forest models implemented in \texttt{scikit-learn}~\cite{scikit}. The quantities analysed were the averaged first transmission-step value and the energy band gap extracted from the ensemble-averaged transmission spectra, as described below. Data extraction, transmission-step detection and descriptor construction were implemented in Python. The \texttt{ruptures} package~\cite{ruptures} was used for transmission-step detection.

\subsection{Automated data preparation}

The large number of transmission spectra made manual identification of the first transmission steps and the energy band gap impractical. We therefore developed an automated procedure to extract these quantities from the ensemble-averaged spectra. 

The first transmission step denotes the first identified step-like region below or above the Fermi level. Its averaged value is the mean ensemble-averaged transmission over the corresponding energy interval. Representative intervals are illustrated in Figs.~\ref{fig:system}(c) and \ref{fig:Tmetallic}(c). Due to the high level of noise observed in the transmission spectra at high temperatures, as well as the frequent occurrence of several peaks at the step boundaries, direct step detection was challenging. Each spectrum was divided at its midpoint, corresponding to the Fermi level ($E_{\mathrm F}=0$), and the two energy ranges were analysed independently, proceeding from the Fermi level towards lower and higher energies. Change-point detection was performed using the Pruned Exact Linear Time (PELT) algorithm~\cite{killick2012optimal} with a radial basis function (RBF) kernel cost function, as implemented in \texttt{ruptures}~\cite{truong2020selective}. The algorithm determines the optimal change points by minimising the sum of segment costs together with a penalty for additional change points. The RBF cost model enables the detection of changes not only in the signal mean but also in its variance and overall distribution. This approach allows reliable identification of the first transmission step even in spectra with high noise.

The energy band gap was determined as the energy separation between the detected transmission steps, while also accounting for cases in which the transmission between the steps was non-zero.

\section{Results and discussion}

To establish relationships between junction structure and electronic transport, including its response to temperature and an external magnetic field, we analysed two complementary transport descriptors across the CNT--CNT junction library: the averaged first transmission-step value and the energy band gap. The first characterises the efficiency of electron transmission when the first set of propagating CNT subbands becomes available, whereas the second describes the energy range around the Fermi level in which no states are available to contribute to transmission. To determine which structural and external parameters are most strongly associated with these quantities, we first used ML analysis to rank the input descriptors by importance and then examined the resulting trends as functions of junction structure, temperature, and perpendicular magnetic field.

\subsection{Averaged first transmission-step}


For an isolated CNT, the energy-dependent transmission has a step-like form, with additional transmission channels becoming available as successive CNT subbands are reached. In a CNT--CNT junction, electrons must additionally tunnel between the two nanotubes. Intertube coupling and coherent interference therefore strongly modulate these ideal transmission plateaux, producing energy-dependent oscillations and resonances even within the energy range associated with the first set of propagating subbands  ~\cite{manuscriptAGA,Taming}. As a result, the transmission of a given junction evaluated at a single energy can depend strongly on the exact position of the electrochemical potential.  
We therefore use the averaged first transmission-step value as a characteristic transport descriptor of each junction. The first transmission step corresponds to the first step-like region encountered when moving away from the Fermi level towards lower or higher energies, where the first propagating CNT subbands contribute to transport. In the spin-degenerate convention used here, the corresponding first plateau of an ideal isolated CNT has a transmission of up to $T=2$ before additional subbands become available. In a CNT--CNT junction this plateau is strongly modulated, but its remnants can still be identified in the transmission spectrum, as illustrated in Fig.~\ref{fig:system}(c) and Fig.~\ref{fig:Tmetallic} (c). The step boundaries were determined using the automated procedure described in the Methods.
For each temperature and magnetic field, the transmission spectrum entering this analysis is already an ensemble average over the 12 thermally perturbed MD realisations. The averaged first transmission-step value is then obtained by averaging this spectrum over the energy interval corresponding to the first step. This procedure reduces the sensitivity of the comparison to a particular choice of electrochemical potential, which is important because the local doping level of individual junctions in an experimental CNT assembly is generally unknown and may vary throughout the sample. The resulting quantity should therefore be regarded as a conductance-related descriptor over a limited range of possible electrochemical potentials rather than as the finite-temperature conductance at a specific doping level.

CNT assemblies are commonly exposed to acids during processing or intentional chemical doping and are therefore typically p-doped~\cite{hayashi2020,manuscriptAGA}. We consequently focus primarily on the first transmission step below the Fermi level, corresponding to the p-doped side of the spectrum, as marked in Fig.~\ref{fig:system}(c) and Fig.~\ref{fig:Tmetallic} (c). For semiconducting junctions, the band-gap region is excluded and the first non-zero transmission step below the gap is considered. The descriptor therefore characterises transport after the relevant CNT states have become accessible and should not be interpreted as the transmission of an undoped semiconducting junction.

\begin{figure*}[h!]
     \centering
     \includegraphics[width=0.99\linewidth]{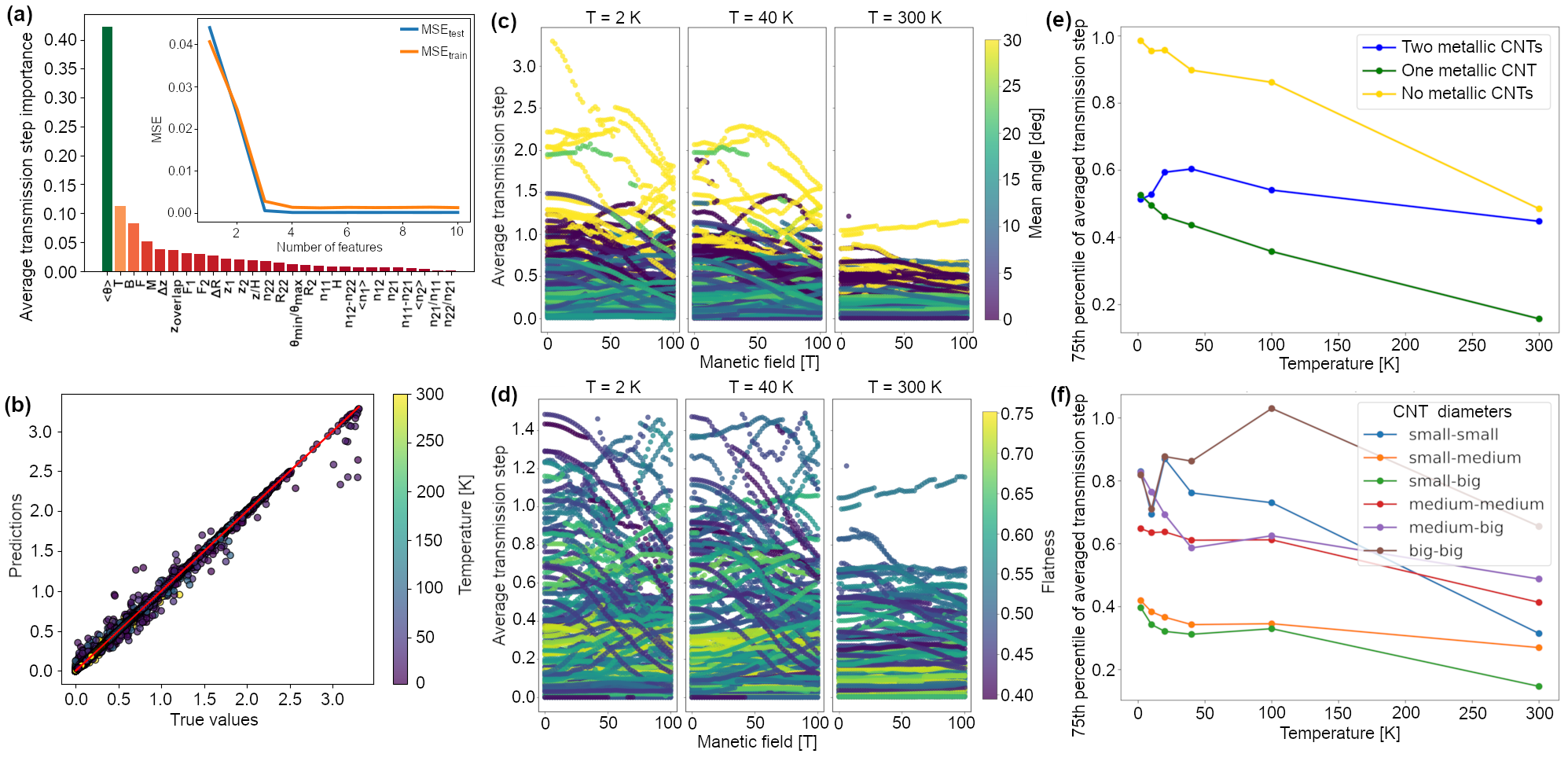}
     \caption{(a) Random-forest analysis of the averaged first transmission-step value for all considered CNT--CNT junctions. Relative importance of the input descriptors, ordered by decreasing importance. The inset shows the training and test mean-square errors (MSEs) as functions of the number of features. (b) Model predictions versus the corresponding values obtained from the MD-TB-Peierls-NEGF calculations, colour-coded by temperature. The red diagonal denotes perfect agreement. (c,d) Averaged first transmission-step value as a function of perpendicular magnetic field at $T=2$, $40$, and $300~\mathrm{K}$, colour-coded by (c) the mean chiral angle of the nanotubes forming the junction and (d) the junction flatness. (e,f) Temperature dependence of the 75th percentile of the averaged first transmission-step value, grouped by (e) junction metallicity and (f) the diameter classes of the two nanotubes forming the junction. The diameter classes are defined for each nanotube as small for $2R < 1~\mathrm{nm}$, medium for $1~\mathrm{nm} \leq 2R < 2~\mathrm{nm}$, and big for $2R \geq 2~\mathrm{nm}$.  }
     \label{fig:steps}
 \end{figure*}

Figure~\ref{fig:steps}(a) shows the relative importance of the input descriptors in the random-forest analysis of the averaged first transmission-step value. The mean chiral angle $\langle\theta\rangle$ is clearly the dominant descriptor, with an importance substantially larger than that of any other individual parameter. Temperature and magnetic field are the next most important descriptors, followed by metallicity and several geometrical parameters. The inset shows that both the training and test mean-square errors decrease sharply when the number of retained descriptors is increased from two to three, whereas adding further descriptors produces only a small additional improvement. This indicates that most of the predictive information is already contained in a small subset of descriptors, consistent with the dominant importance of $\langle\theta\rangle$, temperature, and magnetic field.

The quality of the random-forest description is further shown in Fig.~\ref{fig:steps}(b). Most predicted values lie close to the diagonal corresponding to perfect agreement with the values obtained from the MD--TB--Peierls--NEGF calculations. The largest deviations occur mainly for the highest transmission-step values, which form a comparatively sparse part of the dataset. The model nevertheless reproduces the overall distribution over the complete temperature range.

The dominant role of the mean chiral angle is directly visible in Fig.~\ref{fig:steps}(c), where the averaged first transmission-step value is plotted as a function of perpendicular magnetic field and colour-coded by $\langle\theta\rangle$. Increasing temperature generally compresses the distribution towards lower transmission values, while the magnetic-field dependence remains strongly junction-specific and frequently non-monotonic, consistent with the interference-driven response identified previously~\cite{Taming}. Importantly, signatures of this response remain visible even at $300~\mathrm{K}$. This supports and extends the conclusion of Ref.~\cite{Taming} that interference is not merely a low-temperature effect, but remains relevant for understanding and designing high-transmission CNT junctions even at room temperature.
As can be seen in  Fig.~\ref{fig:steps}(c), the highest transmission-step values are predominantly associated with $\langle\theta\rangle$ close to $30^{\circ}$, corresponding to armchair--armchair junctions. Ideal armchair $(n,n)$ CNTs remain gapless under curvature, with bands crossing at the Fermi level, whereas non-armchair nanotubes satisfying $(n-m)=3k$ can develop small curvature-induced gaps. This provides a favourable electronic basis for high junction transmission, as illustrated in Fig.~\ref{fig:Tmetallic} for the representative $(20,20)+(20,20)$ junction formed by two identical armchair nanotubes. In the present library, the two CNTs forming each junction are parallel. Identical armchair pairs therefore combine metallic constituent nanotubes with matched axial periodicity, avoiding one important source of structural mismatch at the junction. Nevertheless, this does not guarantee high transmission: as shown previously~\cite{Taming}, even parallel armchair--armchair junctions can exhibit strong transmission suppression depending on overlap length and the resulting interference conditions. 

Among the remaining structural descriptors, the junction flatness parameter $F$ also shows a noticeable importance (Fig.~\ref{fig:steps}(a)). This parameter measures the deformation of the junction cross-section from a circular shape and is defined as F=R$_{\textrm{min}}$/$R_{\textrm{max}}$. Thus, F approaches unity for a nearly circular cross-section and decreases as the junction becomes more flattened. To examine this trend more closely, Fig.~\ref{fig:steps}(d) shows the averaged first transmission-step value as a function of magnetic field, colour-coded by F. In order to resolve the behaviour in the low-transmission region, the plotted range is restricted to averaged transmission-step values below 1.5. It becomes clear that larger values of F, corresponding to less flattened junctions, are preferentially associated with lower transmission-step values. 
More strongly flattened junctions, with smaller F, extend over a broader range and can reach  considerably higher transmission values. This trend is consistent with the idea that flattening of the nanotubes in the contact region can enhance intertube coupling by increasing the extent of close contact between the two CNT surfaces~\cite{Yoon2001,Durrant2022}. The association becomes less pronounced at higher temperature. As illustrated in Fig.~\ref{fig:Tmetallic}(b), thermal fluctuations introduce irregular local distortions without producing substantial cross-sectional flattening. The single flatness parameter therefore becomes less representative of the structural changes that influence intertube coupling and transmission at elevated temperature.

Although junction metallicity has a considerably smaller random-forest importance than the mean chiral angle (Fig.~\ref{fig:steps}(a)), it provides a physically natural classification of the junctions and is directly relevant to experimentally available metallicity-enriched CNT samples. Figure~\ref{fig:StepMetallicity} therefore separates the junctions into those formed by two metallic CNTs, one metallic and one semiconducting CNT, and two semiconducting CNTs. The red dashed line in each panel denotes the 75th percentile of the averaged first transmission-step values. We use this percentile as a robust measure of the upper part of the distribution: 75\% of the values lie below this threshold, while it is considerably less sensitive to isolated extreme values than the maximum. The three electronic classes exhibit distinct temperature dependences. For metallic--metallic junctions, the 75th percentile increases slightly between $2$ and $20$--$40~\mathrm{K}$ before decreasing at higher temperatures. In contrast, both mixed metallic--semiconducting and semiconducting--semiconducting junctions show a decrease with increasing temperature. These trends are summarised in Fig.~\ref{fig:steps}(e), where the magnetic-field dependence has been integrated into the statistical distribution by pooling the field-dependent values at each temperature. An important result is that semiconducting--semiconducting junctions exhibit the highest 75th-percentile transmission-step values over the full temperature range considered. This does not imply that undoped semiconducting junctions are more conductive than metallic ones: at the Fermi level their band gap suppresses transport. Rather, the comparison is made within the first conducting step, corresponding to a sufficiently shifted electrochemical potential. Once semiconducting CNTs are brought into a conducting regime by doping, their available transmission channels can provide larger junction transmission than those of metallic CNTs. This behaviour is consistent with experimental studies of type-separated CNT networks, where redox doping was found to enhance intertube transmission more strongly in semiconductor-enriched than in metal-enriched films \cite{Blackburn2008}, and with recent modelling showing that sufficiently doped semiconducting CNT bundles can exceed comparable metallic bundles in transmission \cite{Bulmer}. At $300~\mathrm{K}$ the difference between semiconducting--semiconducting and metallic--metallic junctions is strongly reduced, but the former still retain a slightly higher 75th-percentile transmission-step value. Mixed metallic--semiconducting junctions show the lowest value at this temperature.
 
We next considered nanotube diameter, another experimentally accessible and physically intuitive structural classification. Figures~\ref{fig:StepSizesAngle} and \ref{fig:StepSizesMetallicity} show the field dependence of the averaged first transmission-step value for the six possible combinations of small, medium, and large nanotubes, colour-coded by the mean chiral angle and junction metallicity, respectively.Consistent with the relatively low importance of the diameter-related descriptors in Fig.~\ref{fig:steps}(a), the differences between these classes are less pronounced than those associated with chiral angle. Nevertheless, clear trends remain, as summarised by the 75th percentiles in Fig.~\ref{fig:steps}(f). Small--medium and small--big junctions consistently show the lowest values, whereas big--big junctions exhibit the highest 75th-percentile transmission from approximately $20~\mathrm{K}$ onwards. Small--small junctions also show comparatively high transmission at low temperature. This suggests that both nanotube size and diameter matching influence junction transport, with strong diameter mismatch involving a small CNT being particularly unfavourable. Two geometrical effects may contribute to this trend. Larger nanotubes have smaller surface curvature and provide a broader near-contact region at a fixed intertube separation, increasing the number of atomic pairs that can contribute to intertube coupling. At the same time, a strong diameter mismatch can reduce structural and electronic matching across the junction. These effects provide a plausible explanation for the relatively low transmission of small--medium and small--large junctions. The non-monotonic temperature dependence observed for some diameter-matched classes, most notably the large--large junctions, further shows that diameter alone does not determine transport. Because junction transmission is interference- and registry-dependent, moderate thermal distortions can move a junction either towards or away from a favourable coupling configuration, whereas stronger thermal disorder at high temperature generally suppresses the transmission-step values.

Finally, we compared the first transmission steps on the p- and n-doped sides of the Fermi level. Figure~\ref{fig:StepLR} shows the difference between the averaged first transmission-step values below and above $E_{\mathrm F}$ as a function of magnetic field. The two sides of the spectrum are clearly not equivalent, and the asymmetry depends strongly on nanotube chirality. Such electron--hole asymmetry is physically expected in realistic CNT junctions. Exact electron--hole symmetry is a special property of the ideal orthogonal nearest-neighbour $\pi$-orbital model and is broken when more realistic effects, including orbital overlap and junction-specific intertube coupling, are taken into account~\cite{Li2006}. Consequently, valence- and conduction-side states need not couple equivalently across the junction and can give different transmission. Increasing temperature substantially reduces the magnitude of the asymmetry, although a finite difference between the two sides remains visible even at $300~\mathrm{K}$.

\subsection{Energy band gap}

\begin{figure*}[h!]
     \centering
     \includegraphics[width=0.99\linewidth]{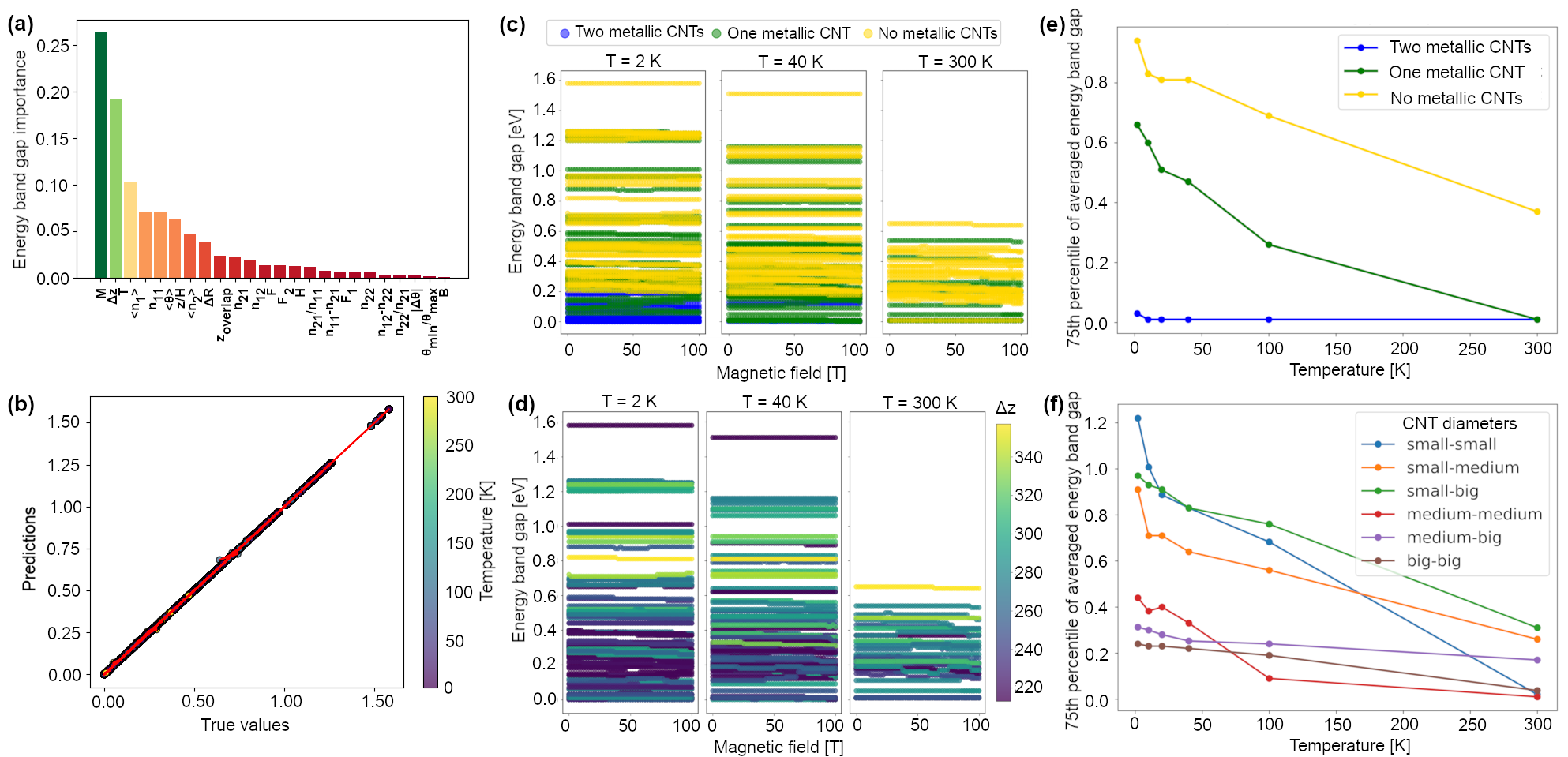}
     \caption{(a) Random-forest analysis of the energy band gap for all considered CNT--CNT junctions. Relative importance of the input descriptors, ordered by decreasing importance. (b) Model predictions versus the corresponding values obtained from the MD--TB--Peierls--NEGF calculations, colour-coded by temperature. The red diagonal denotes perfect agreement.  (c,d) Energy band gap as a function of perpendicular magnetic field at $\mathrm{T}=2$, $40$, and $300~\mathrm{K}$, colour-coded by (c) the number of metallic nanotubes forming the junction and (d) the difference in nanotube lengths, $\Delta z$. (e,f) Temperature dependence of the 75th percentile of the energy band gap, grouped by (e) junction metallicity and (f) the diameter classes of the two nanotubes forming the junction.}
     \label{fig:gaps}
 \end{figure*}

We next analysed the energy band gap extracted from the ensemble-averaged transmission spectra. Importantly, the gap is a property of the complete junction and is not determined solely by whether the constituent nanotubes are metallic or semiconducting. Even a junction formed by two metallic CNTs can develop a finite gap or a strongly suppressed transmission region around the Fermi level as a result of intertube coupling, hybridisation, structural mismatch, and quantum interference. In particular, our previous work showed coupling-induced gap opening and transmission quenching even in parallel armchair--armchair junctions for appropriate overlap lengths~\cite{manuscriptAGA,Taming}.

The random-forest analysis in Fig.~\ref{fig:gaps}(a) reveals a markedly different hierarchy of input descriptors from that obtained for the averaged first transmission-step value. Junction metallicity is now clearly the dominant descriptor, which is expected because the availability of states around the Fermi level is directly related to the electronic character of the constituent nanotubes. The difference in nanotube lengths, $\Delta z$, is the second most important descriptor, followed by several parameters related to nanotube chirality and junction geometry. In contrast, magnetic field has only a very small importance. Figure~\ref{fig:gaps}(b) shows that the resulting model reproduces the calculated gap values closely over the full range considered, with the predictions lying near the line of perfect agreement.

The dominant role of metallicity is directly visible in Fig.~\ref{fig:gaps}(c). The energy gap shows only a weak dependence on perpendicular magnetic field, consistent with the low importance of B in Fig.~\ref{fig:gaps}(a), whereas the separation according to junction metallicity is pronounced. Junctions formed by two semiconducting CNTs generally exhibit the largest gaps, mixed metallic--semiconducting junctions show intermediate values, and metallic--metallic junctions are concentrated close to zero, although finite gaps are also present in some metallic--metallic systems. Increasing temperature reduces both the magnitude and the spread  of the extracted gaps. Within the present ensemble-averaged transmission analysis, this can be understood as a consequence of thermally induced structural fluctuations, which introduce finite transmission into energy regions that are more strongly suppressed at low temperature and therefore reduce the detected gap. This should not necessarily be interpreted as a conventional intrinsic band-gap renormalisation of an isolated CNT. 
The metallicity dependence is shown in more detail in Fig.~\ref{fig:BGMetallicity} and summarised by the 75th percentiles in Fig.~\ref{fig:gaps}(e). The gap decreases with temperature in all three junction classes, but at markedly different rates. For metallic--metallic junctions the 75th-percentile value is already close to zero at low temperature. For mixed junctions it decreases from approximately $0.66~\mathrm{eV}$ at $2~\mathrm{K}$ to approximately $0.01~\mathrm{eV}$ at $300~\mathrm{K}$. In contrast, semiconducting--semiconducting junctions retain a substantial gap even at room temperature, with a 75th-percentile value of approximately $0.37~\mathrm{eV}$.

The relatively high importance of $\Delta z$ requires more careful interpretation. Figure~\ref{fig:gaps}(d) does not show a simple monotonic relationship between the energy gap and $\Delta z$, indicating that its importance in the random-forest model does not correspond to a direct one-parameter dependence. In the present library, nanotubes with different chiralities have different translational periods, and the target overlap length of approximately $20~\mathrm{nm}$ could therefore not be reproduced identically for every junction. As a result, $\Delta z$ is linked to other geometrical properties of the constructed junctions, including the realised overlap length. 
The overlap length itself has a clear physical connection to junction transport because it determines the region over which intertube tunnelling occurs and therefore controls the hybridisation and relative phases of the electronic paths through the junction~\cite{manuscriptAGA,Taming}. Even small changes in overlap can consequently produce substantial changes in transmission and gap opening. Figure~\ref{fig:BGAngle} shows that, even within the relatively narrow range of overlap lengths sampled in the present library, a broad distribution of energy gaps is obtained, particularly at low temperature. No simple one-to-one dependence is observed, however, because the effect of overlap is strongly coupled to nanotube chirality, metallicity, and atomic registry.

Finally, we considered the nanotube diameter classes (Fig.~\ref{fig:BGSizes}). At low and intermediate temperatures, junctions containing smaller CNTs generally exhibit larger energy
gaps. This is particularly clear at $2~\mathrm{K}$, where the 75th percentile decreases from approximately $1.22~\mathrm{eV}$ for small--small junctions to $0.44~\mathrm{eV}$ for
medium--medium and $0.24~\mathrm{eV}$ for big--big junctions (Fig.~\ref{fig:gaps}(f)). The same general ordering persists over much of the temperature range, although it becomes progressively weaker as the gaps decrease. At $300~\mathrm{K}$ the ordering is no longer monotonic: the gap distribution of the small--small class collapses almost completely, whereas small--medium and small--big junctions retain larger 75th-percentile values. Thus, although smaller nanotube diameters tend to favour larger gaps at low temperature, diameter alone cannot determine the junction gap. This is consistent with the dominant role of metallicity identified in Fig.~\ref{fig:gaps}(a), since each diameter class contains a different mixture of metallicities and chiralities.

\section*{Conclusions}

By analysing a structurally diverse library of 146 SWCNT--SWCNT junctions, we move from the behaviour of individual representative systems to statistical structure--transport relationships across a broad range of nanotube chiralities. The two transport characteristics considered here are governed by distinctly different hierarchies. The averaged first transmission-step value is controlled primarily by the mean chiral angle of the two nanotubes forming the junction, whereas the energy gap is governed predominantly by the metallic or semiconducting character of the constituent CNTs. Thus, the factors that determine how efficiently electrons are transferred across a junction once propagating states are available are not the same as those controlling whether such states are available near the Fermi level. Temperature also acts differently on the two quantities: thermal structural fluctuations generally suppress the averaged transmission while simultaneously reducing the extracted energy gap. The magnetic field, in contrast, has a much stronger influence on transmission than on the gap, and signatures of the interference-driven field response remain visible even at $300~\mathrm{K}$. 
Quantum interference in CNT junctions should therefore not be regarded solely as a cryogenic phenomenon, but as a mechanism that remains relevant for understanding and designing conductive CNT assemblies under technologically relevant conditions. The distinction between the availability of conducting states and the efficiency of their transmission is particularly evident for semiconducting--semiconducting junctions. These retain the largest gaps, but once shifted into their conducting regime they can exhibit transmission comparable to or higher than that of metallic junctions. Metallic character alone is therefore not sufficient to infer the transport quality of an individual
CNT--CNT contact. 
These findings provide a basis for future network-scale models that account explicitly for the diversity of CNT junctions.

\section*{Author contributions}
K.Z.M. conceived the study and supervised the project. J.A.C.M. developed the automated computational workflow and the tight-binding implementation. T.K. contributed to the \texttt{sisl} implementation, and I.V.L. contributed to the \texttt{TBtrans} implementation.
J.A.C.M. and T.K. performed the numerical calculations under  supervision of K.Z.M., who also contributed to the calculations. M.M. performed the machine-learning analysis. K.Z.M. prepared the first draft of the manuscript with contributions from J.A.C.M. and T.K.. K.Z.M., I.V.L., T.K. and J.A.M. contributed to the interpretation of the theoretical results. All authors discussed the results, reviewed the manuscript and contributed to its final version.

\section*{Conflicts of interest}
There are no conflicts to declare.

\section*{Data availability}
The data supporting the findings of this study, including the processed transmission spectra, structural models, and derived transport descriptors used in the statistical and machine-learning analyses, are available from the corresponding authors upon reasonable request. The modified versions of \texttt{sisl} and \texttt{TBtrans} used for the magnetic-field transport calculations are also available from the corresponding authors upon reasonable request, subject to the licences of the original \texttt{sisl} and \texttt{TBtrans} packages. Additional structural models, transmission spectra, and detailed analyses supporting the results are provided in the Supplementary Information (SI), available at DOI:...

\section*{Acknowledgements}
The authors thank Magdalena Marganska for helpful discussions on the tight-binding parametrisation and the construction of the intra-nanotube Hamiltonian. 
J.A.C.M., T.K., I.V.L. and K.Z.M acknowledge the technical and human support provided by the DIPC Supercomputing Center, Spain. K.Z.M. is grateful to the Agencia Estatal de Investigación, Ministerio de Ciencia e Innovación, Spain, for funding this research through the Proyectos de Generación de Conocimiento 2022 programme (PID2022-139776NB-C65) and the Proyectos de Generación de Conocimiento 2025 programme (PID2025-174822NB-C32). K.Z.M also would like to thank the European Commission (Marie Sklodowska-Curie Cofund Programme; grant no. H2020-MSCA-COFUND-2020-101034228-WOLFRAM2) for supporting this research.  K.Z.M also would like to aknowledge the support from the Ramón y Cajal grant RYC2024-051436-I, funded by MICIU/AEI/10.13039/501100011033 and by ESF+. JAM acknowledges the support from Centera2 project (FENG.02.02-IP.02.01-IP.05-T0004/23) funded with IRA FENG program of Foundation for Polish Science, and co-financed by the EU FENG Programme. I.V.L.  acknowledges support from the EuroHPC JU under the MAX (Materials design at the Exascale) project (grant no. 101093374), and from the Spanish MCIN/AEI/10.13039/501100011033 and the European Union NextGenerationEU/PRTR through grant no. PCI2022-134972-2. ICN2 is supported by the CERCA programme (Generalitat de Catalunya) and the Severo Ochoa Centres of Excellence programme (grant no. CEX2021-001214-S), funded by MCIN/AEI/10.13039/501100011033.

\bibliographystyle{rsc}
\bibliography{biblio}

\newpage
\clearpage
\onecolumn
\thispagestyle{plain}

\begin{center}
    \LARGE \textbf{Supporting Information} \\[0.5em]
    
    \Large
    for \\[0.5em]
    
    \LARGE{\bf 
    Library of carbon nanotube junctions: data-driven insights into structure–magnetotransport relationships
    }
\end{center}

\vspace{2cm}

\setcounter{section}{0}
\setcounter{subsection}{0}
\setcounter{figure}{0}
\setcounter{table}{0}

\renewcommand{\thesection}{S\arabic{section}}
\renewcommand{\thesubsection}{S\arabic{section}.\arabic{subsection}}

\renewcommand{\thefigure}{S\arabic{figure}}
\renewcommand{\thetable}{S\arabic{table}}

\appendix

\begin{figure}[h!tb]
    \centering
    \includegraphics[width=0.99\linewidth]{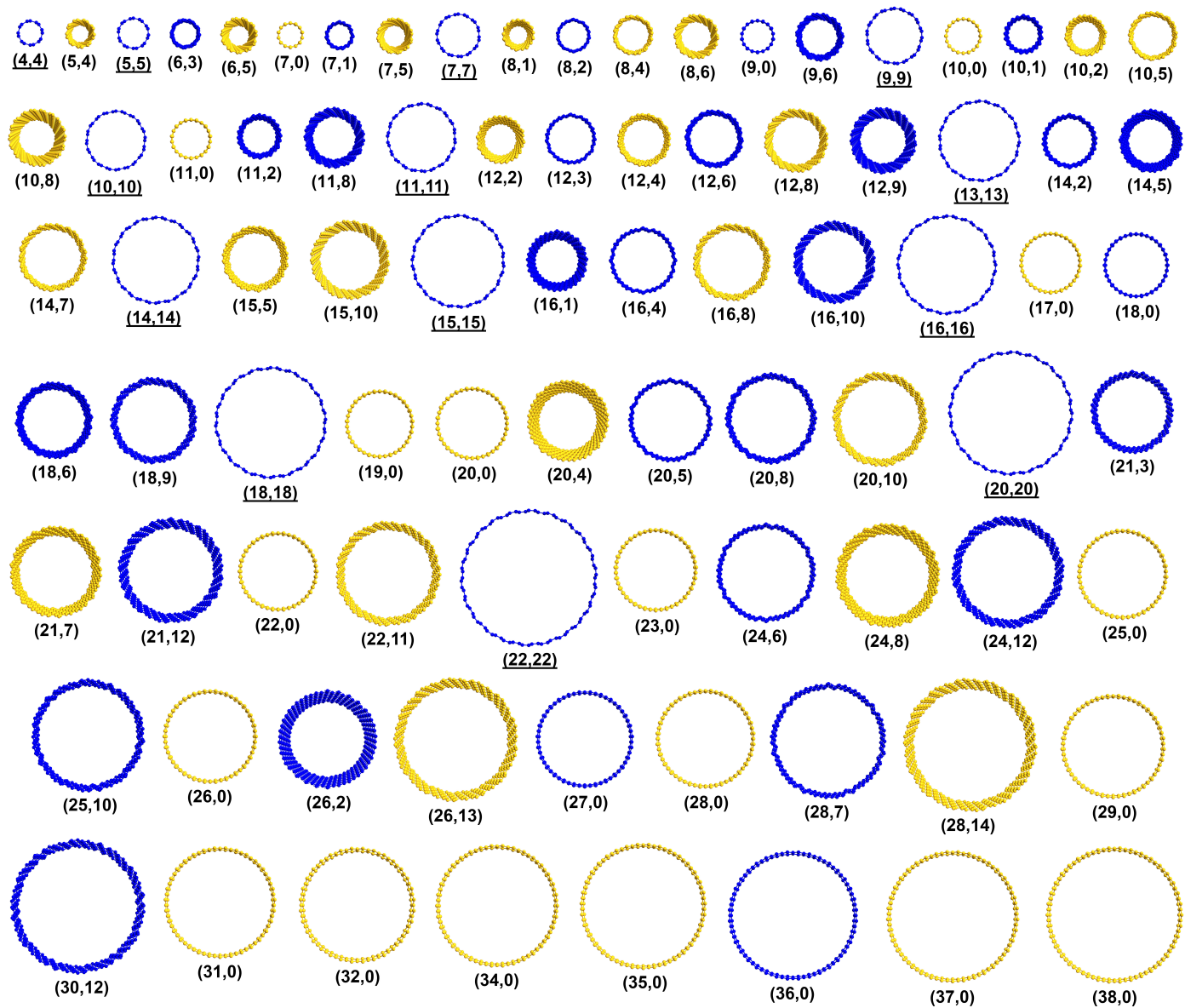}
    \caption{ Perspective views of one translational unit cell of each CNT type used to construct the CNT--CNT junction library. Semiconducting nanotubes are shown in yellow, while nanotubes satisfying the metallicity condition $(n-m)=3k$, with integer $k$, are shown in blue. Armchair nanotubes $(n,n)$ are indicated by underlined chirality labels.}
    \label{fig:CNTs}
\end{figure}

\begin{figure}
     \centering
     \includegraphics[width=1.0\linewidth]{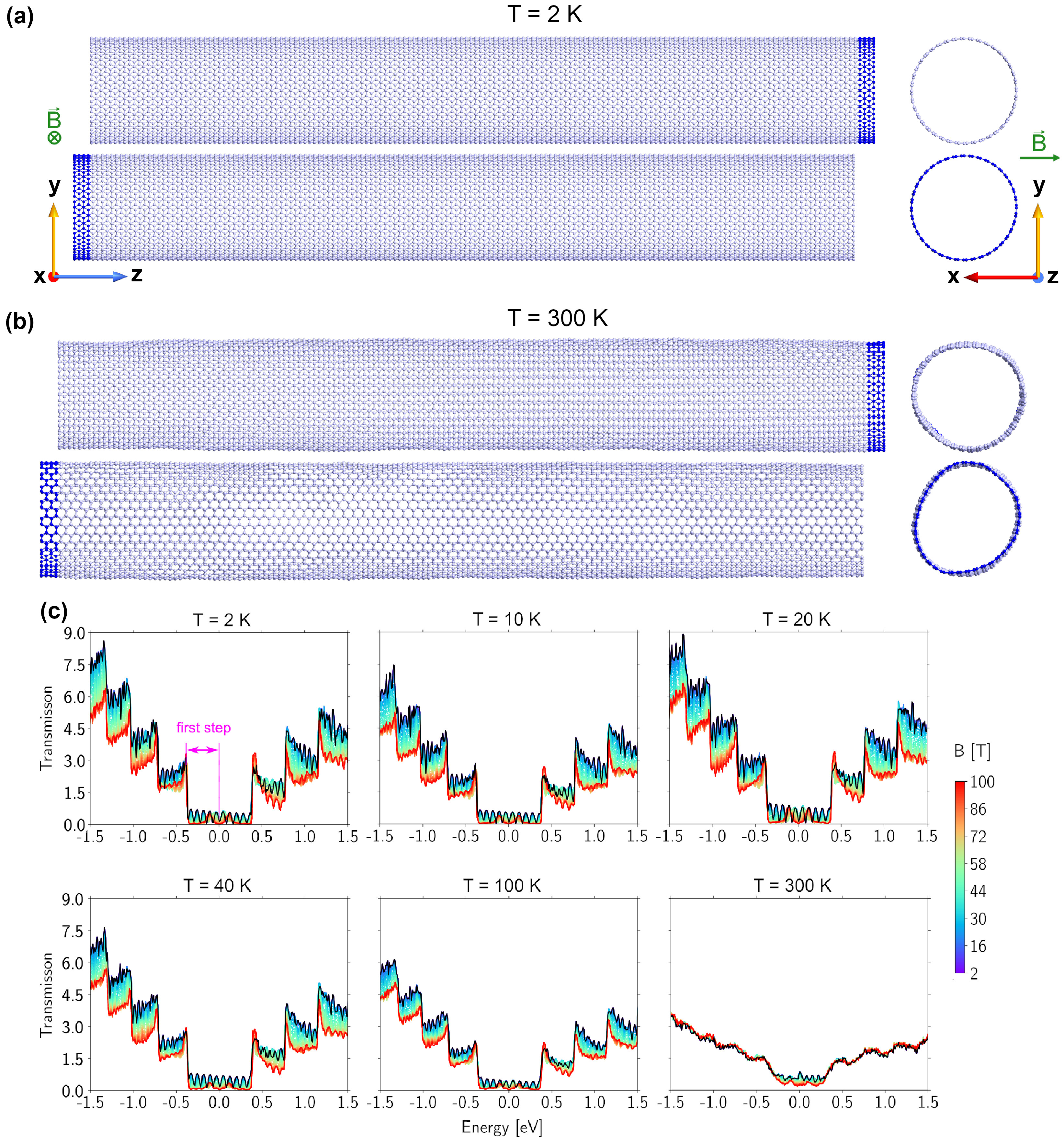}
     \caption{Side and cross-sectional views of the metallic-metallic $(20,20)+(20,20)$ junction at (a) $2~\mathrm{K}$ and (b) $300~\mathrm{K}$. The two metallic $(20,20)$ nanotubes (blue) are connected to left and right semi-infinite electrodes of the same chirality (navy blue). For this junction, $n_{11}=n_{12}=n_{21}=n_{22}=20$ and $\theta_{1}=\theta_{2}=30^{\circ}$. (c) Energy-resolved, ensemble-averaged transmission spectra of the $(20,20)+(20,20)$ CNT junction with an overlap length of $z_{\mathrm{overlap}}=24.58~\mathrm{nm}$. Black lines show the averaged zero-field spectra, whereas coloured lines correspond to averaged spectra under perpendicular magnetic fields up to $100~\mathrm{T}$. The first transmission step on the p-doped side of the spectrum, below the Fermi level, used in the subsequent analysis is indicated for a representative averaged spectrum.
     }
     \label{fig:Tmetallic}
\end{figure}



\begin{figure}[h!tb]
    \centering
    \includegraphics[width=0.75\linewidth]{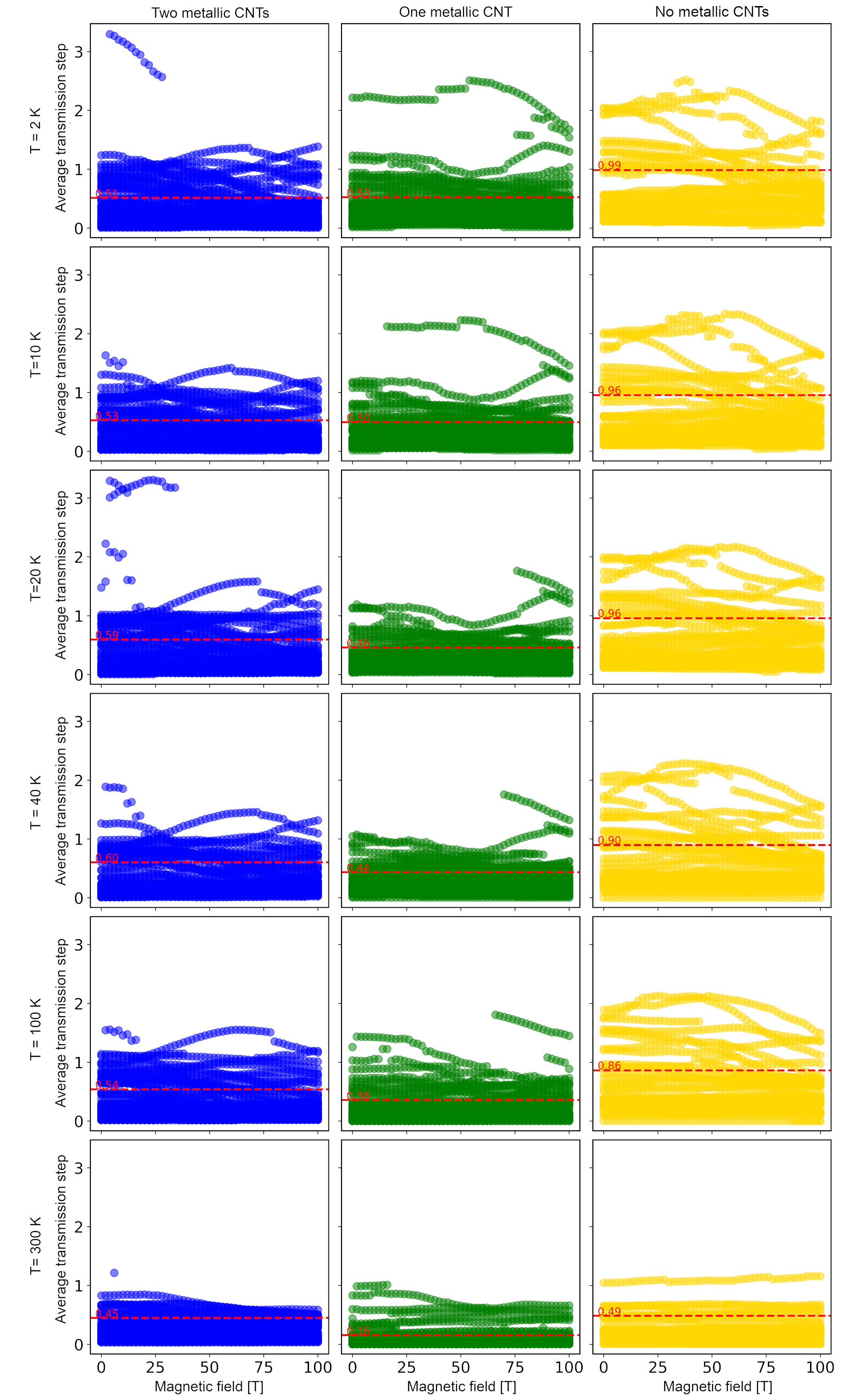}
    \caption{Average transmission step versus magnetic field for different temperatures, grouped by junction metallicity. The red dashed line marks the value below which 75\% of the data points are found.}
    \label{fig:StepMetallicity}
\end{figure}

\begin{figure}[h!tb]
    \centering
    \includegraphics[width=0.75\linewidth]{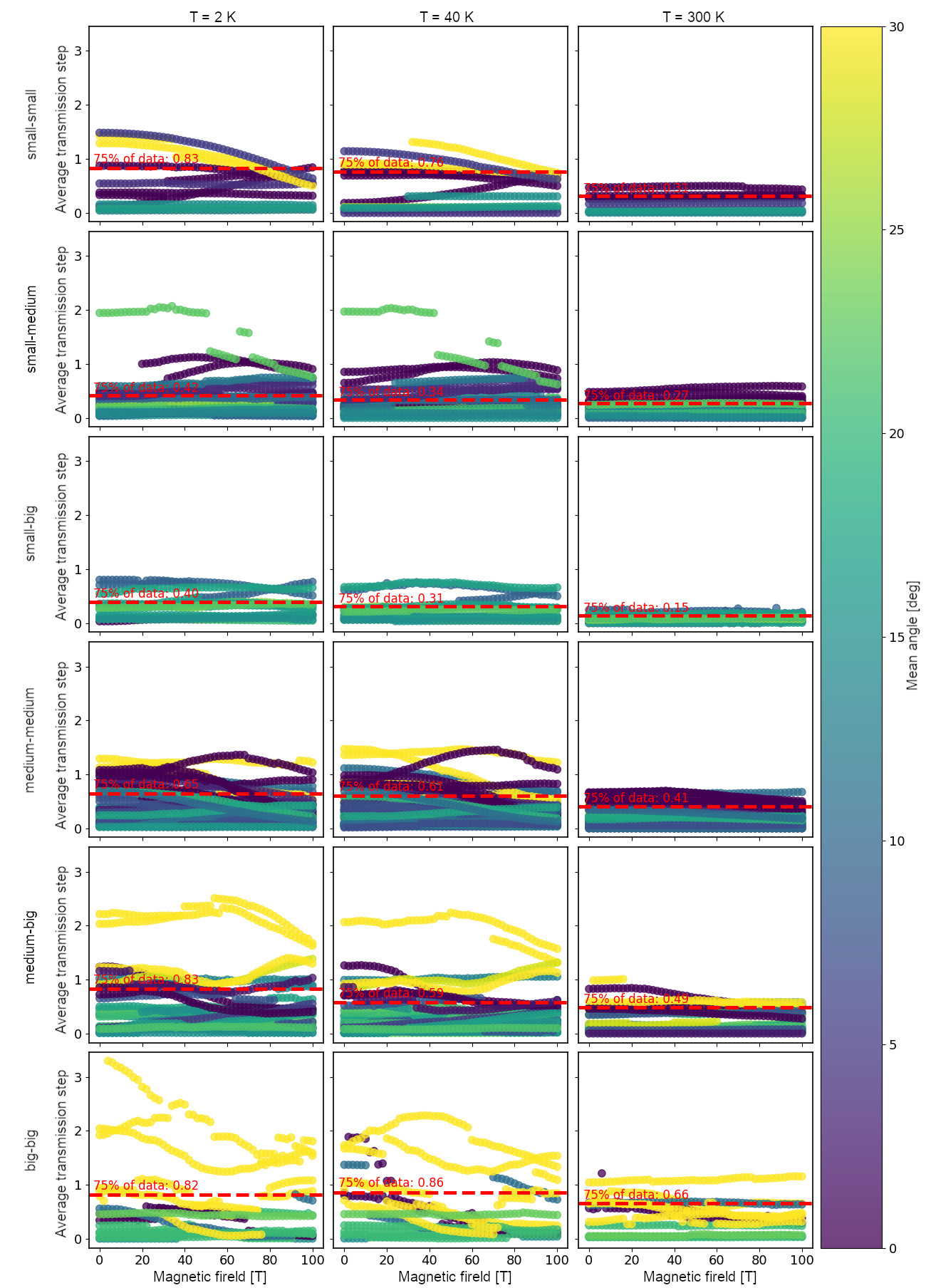}
    \caption{Average transmission step versus magnetic field for different temperatures, grouped by the diameter classes of the two nanotubes forming the junction. The diameter classes are defined for each nanotube as small for $2R < 1~\mathrm{nm}$, medium for $1~\mathrm{nm} \leq 2R < 2~\mathrm{nm}$, and big for $2R \geq 2~\mathrm{nm}$. The colour scale denotes the mean chiral angle of both nanotubes. The red dashed line marks the value below which 75\% of the data points are found.}
    \label{fig:StepSizesAngle}
\end{figure}

\begin{figure}[h!tb]
    \centering
    \includegraphics[width=0.75\linewidth]{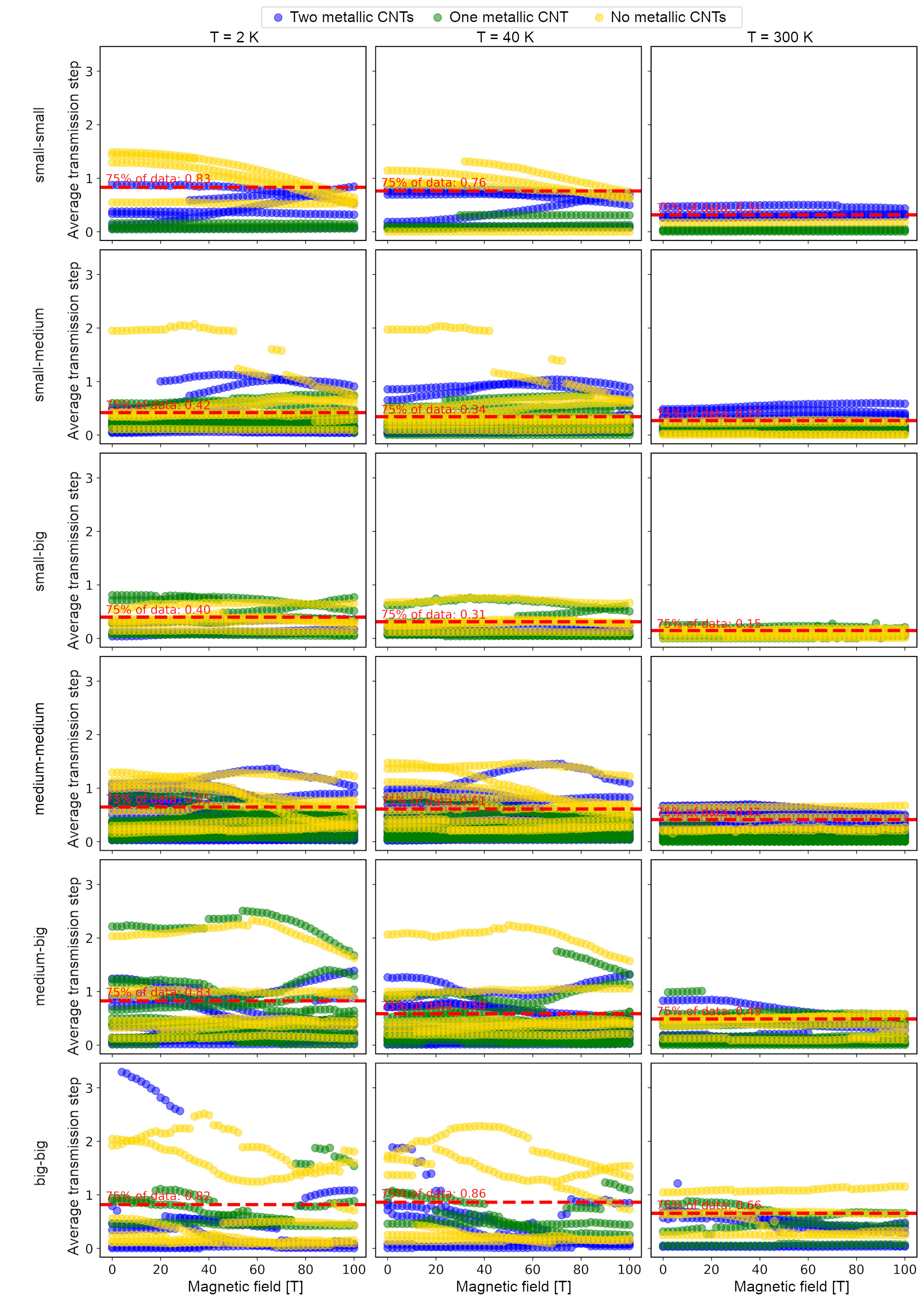}
    \caption{Average transmission step versus magnetic field for different temperatures, grouped by the diameter classes of the two nanotubes forming the junction. The diameter classes are defined for each nanotube as small for $2R < 1~\mathrm{nm}$, medium for $1~\mathrm{nm} \leq 2R < 2~\mathrm{nm}$, and big for $2R \geq 2~\mathrm{nm}$. The colours indicate the metallicity of the junction: blue for metallic--metallic, yellow for semiconducting--semiconducting, and green for metallic--semiconducting CNT pairs. The red dashed line marks the value below which 75\% of the data points are found.}
    \label{fig:StepSizesMetallicity}
\end{figure}

\newpage

\begin{figure}[h!tb]
    \centering
    \includegraphics[width=0.8\linewidth]{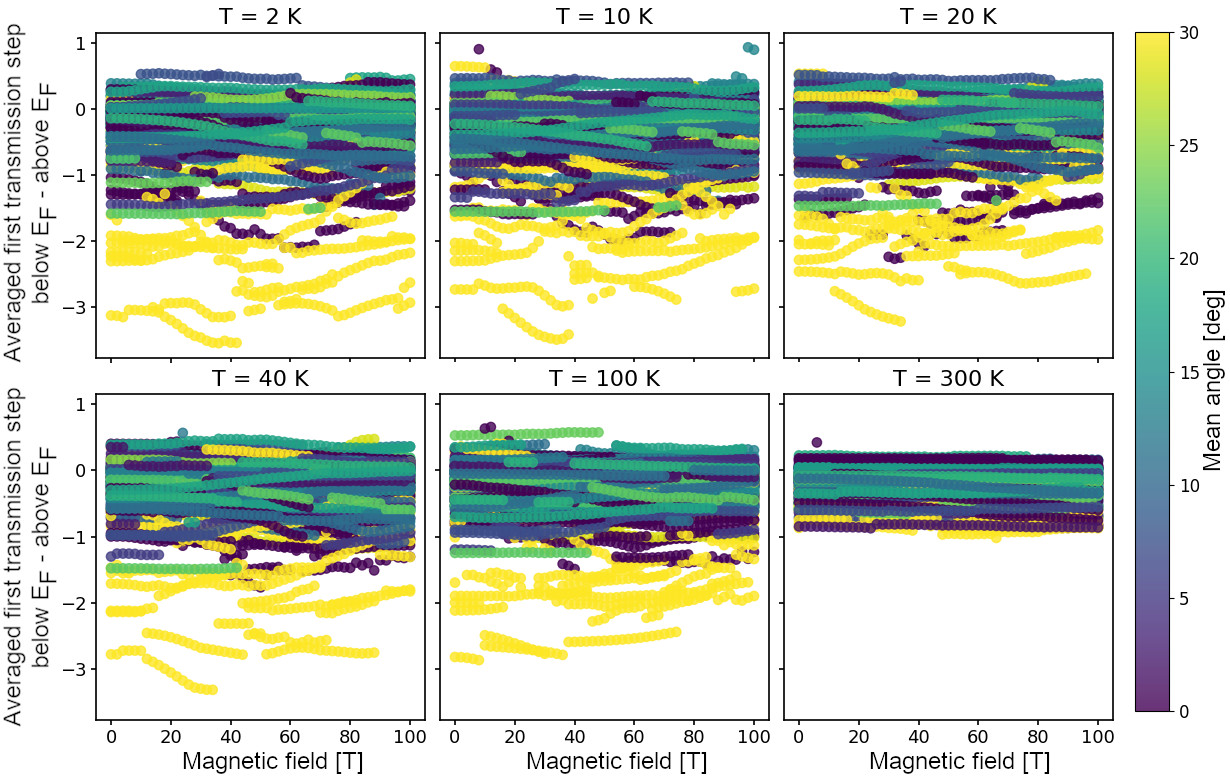}
    \caption{Difference between the averaged first transmission steps below and above the Fermi level, corresponding to the p- and n-doped sides of the spectrum, respectively, as a function of magnetic field. Panels show different temperatures, with colour indicating the mean chiral angle of the junction.}
    \label{fig:StepLR}
\end{figure}


\begin{figure}[h!tb]
    \centering
    \includegraphics[width=0.8\linewidth]{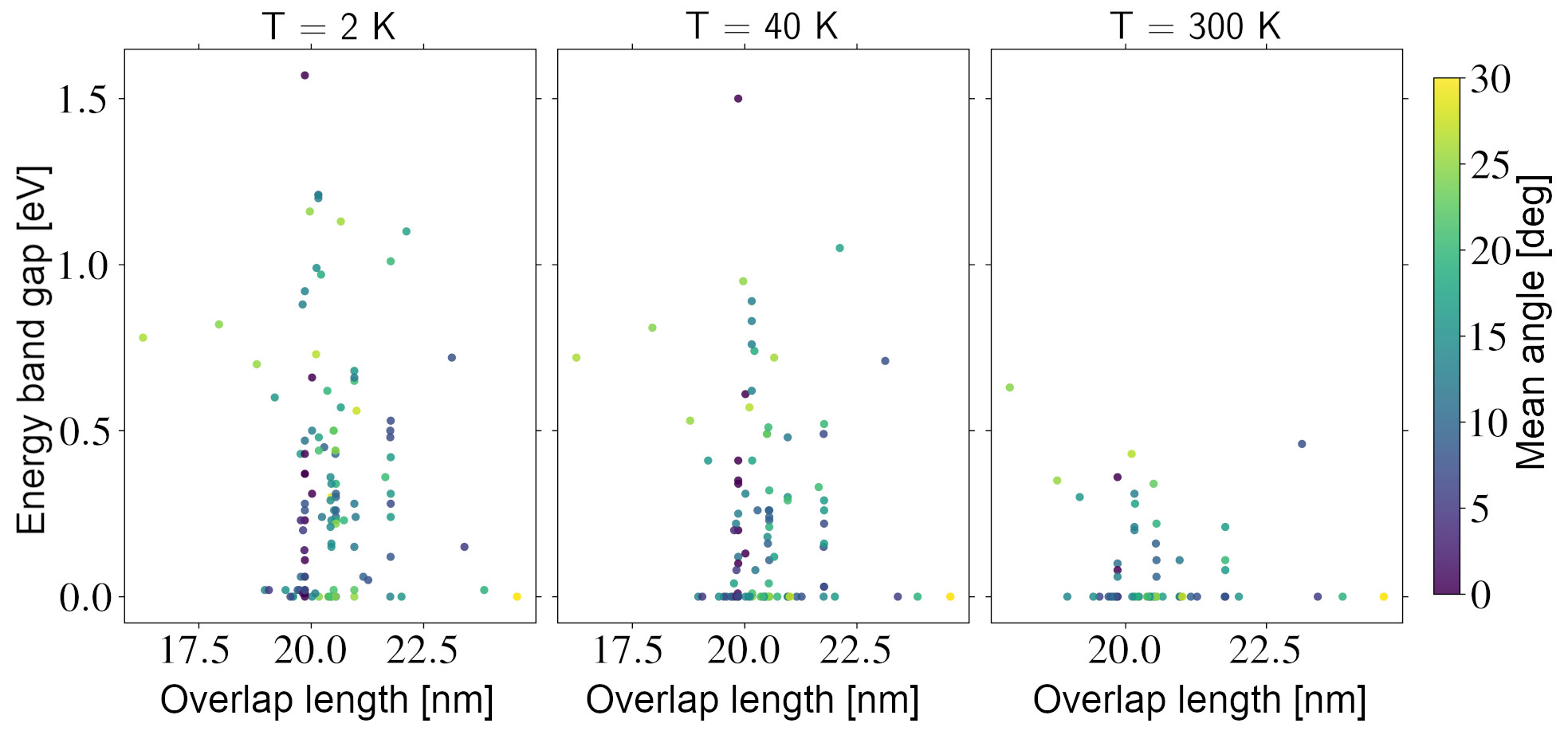}
    \caption{Energy band gap versus overlap length, coloured by mean chiral angle of both nanotubes. }
    \label{fig:BGAngle}
\end{figure}

\begin{figure}[h!tb]
    \centering
    \includegraphics[width=0.7\linewidth]{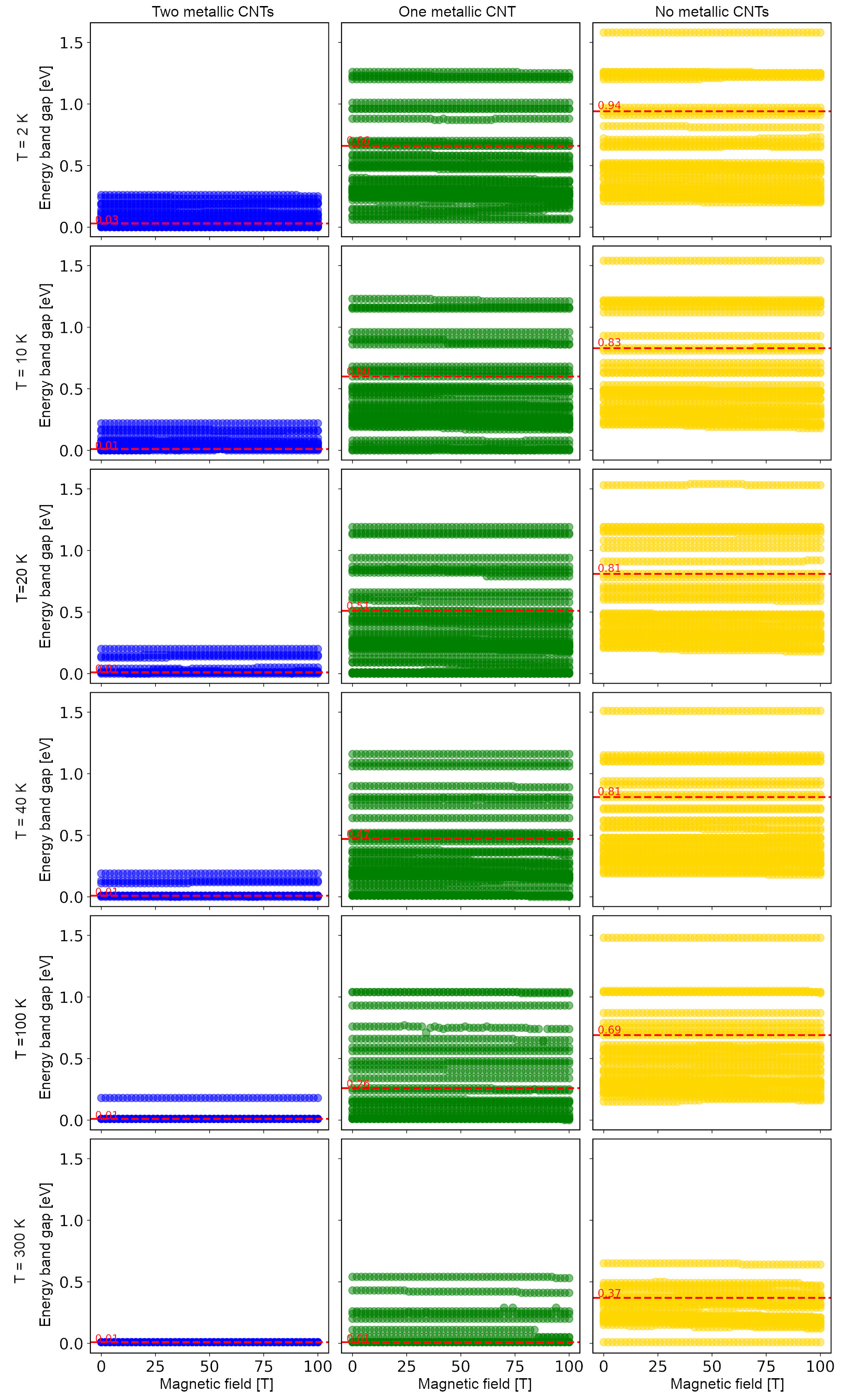}
    \caption{Energy band gap versus magnetic field for different temperatures, grouped by junction metallicity. The red dashed line marks the value below which 75\% of the data points are found.}
    \label{fig:BGMetallicity}
\end{figure}

\begin{figure}[h!tb]
    \centering
    \includegraphics[width=0.8\linewidth]{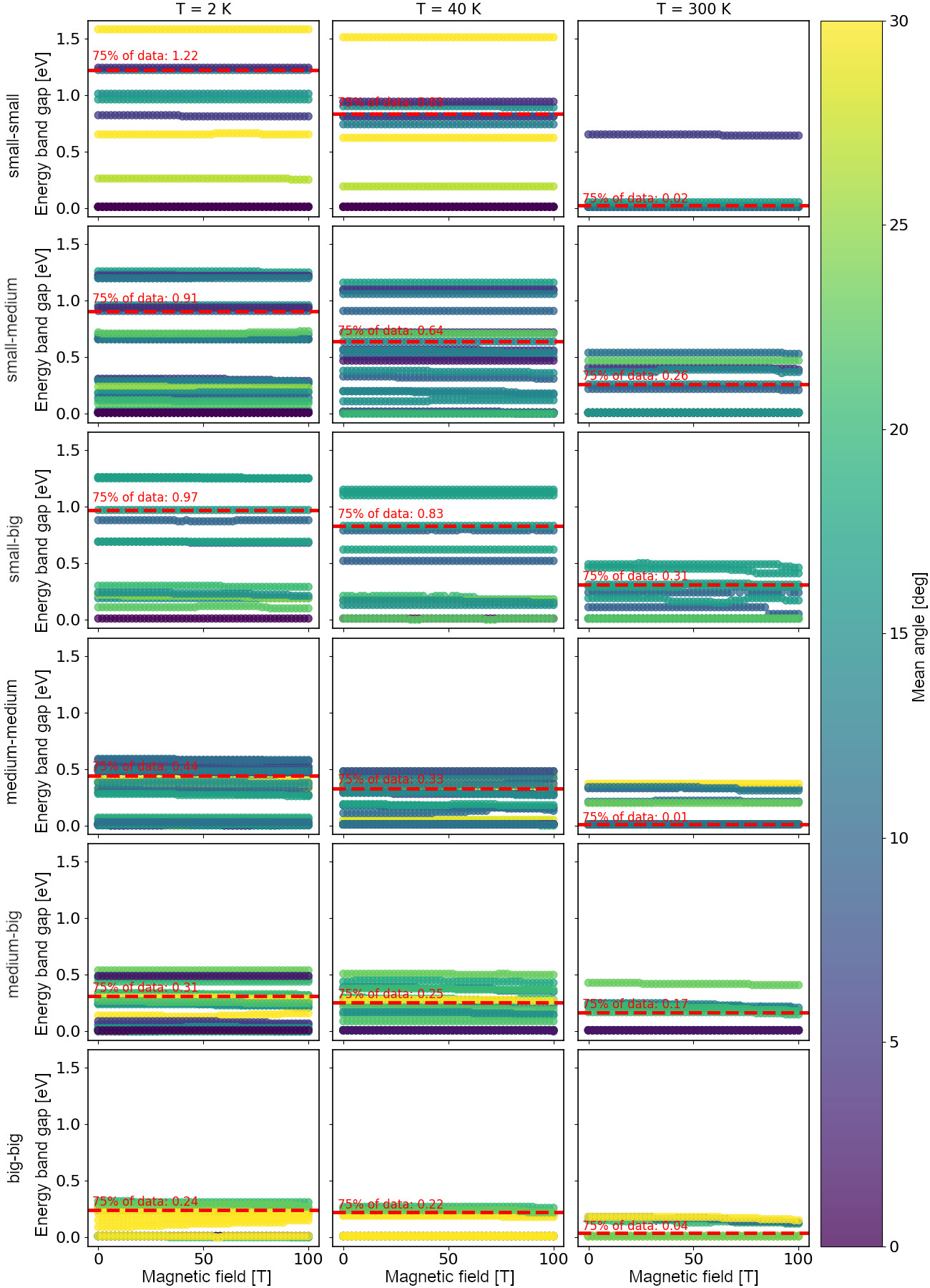}
    \caption{Energy band gap versus magnetic field for different temperatures, grouped by the diameter classes of the two nanotubes forming the junction. The diameter classes are defined for each nanotube as small for $2R < 1~\mathrm{nm}$, medium for $1~\mathrm{nm} \leq 2R < 2~\mathrm{nm}$, and big for $2R \geq 2~\mathrm{nm}$. The colours indicate the metallicity of the junction: blue for metallic--metallic, yellow for semiconducting--semiconducting, and green for metallic--semiconducting CNT pairs. The red dashed line marks the value below which 75\% of the data points are found.}
    \label{fig:BGSizes}
\end{figure}

\newpage
\newpage





\balance


\newpage
\FloatBarrier
\clearpage

\end{document}